\documentclass[aps, prb, twocolumn, showpacs, 10pt, floatfix, superscriptaddress, floatfix, longbibliography]{revtex4-2}
\usepackage{lipsum}  
\usepackage[english]{babel}
\usepackage[utf8]{inputenc}
\usepackage[T1]{fontenc}

\usepackage{amsmath}
\usepackage{braket}
\usepackage{amssymb}
\usepackage[normalem]{ulem}
\usepackage{dsfont}
\usepackage{bm}
\usepackage[dvipsnames]{xcolor}
\usepackage{stmaryrd}
\usepackage{bbm}
\usepackage{mathtools}
\usepackage{hyperref}

\usepackage{graphicx}
\usepackage{dcolumn}
\usepackage{bm}

\hypersetup{
  colorlinks = True,
  citecolor = Maroon,
  linkcolor = Blue,
  urlcolor = Blue,
}

\makeatletter
\newcommand{\thickhline}{%
\noalign {\ifnum 0=`}\fi \hrule height 1.2pt
\futurelet \reserved@a \@xhline%
}
\newcolumntype{"}{@{%
\hskip\tabcolsep\vrule width 1.2pt\hskip\tabcolsep}%
}
\makeatother

\newcommand\Tstrut{\rule{0pt}{2.5ex}}

\newcommand{\wick}[1]{\left. :\! \hspace{-0.5pt} #1 \hspace{-0.5pt} \!: \right.}

\newcommand{\sgn}{\operatorname{sgn}}

\newcommand{\Tr}{\operatorname{Tr}}

\newcommand{\pdag}{^{\vphantom{\dagger}}}
\renewcommand{\Re}{\operatorname{Re}}
\renewcommand{\Im}{\operatorname{Im}}

\newcommand{\dd}{\mathrm{d}}

\newcommand{\SU}{\mathrm{SU}}
\newcommand{\su}{\mathfrak{su}}

\begin{document}

%============================================================================

\title{Chiral instabilities in driven-dissipative quantum liquids}

\author{Zhi-Xing Lin}
\affiliation{Department of Physics, Princeton University, Princeton, New Jersey 08544, USA}

\author{Bastien Lapierre}
\affiliation{Department of Physics, Princeton University, Princeton, New Jersey 08544, USA}

\author{Per Moosavi}
\affiliation{Department of Physics, Stockholm University, 10691 Stockholm, Sweden}

\author{Shinsei Ryu}
\affiliation{Department of Physics, Princeton University, Princeton, New Jersey 08544, USA}

\date{April 28, 2025}

%============================================================================

\begin{abstract}
We investigate the nonequilibrium dynamics of periodically driven Tomonaga-Luttinger liquids (TLLs) coupled to a thermal bath using a Floquet-Lindblad approach. When the coupling to the bath satisfies detailed balance, we obtain a condition for parametric instabilities to be suppressed, symmetrically for both chiralities. Remarkably, by designing a purely chiral coupling to the bath, instead of instability suppression, we uncover a driven-dissipative phase transition between the former symmetric parametric instability and a new chiral parametric instability. In the latter, a single chirality of bosonic quasiparticles gets exponentially amplified, leading to a dynamical chiral imbalance within the TLL, reminiscent of the non-Hermitian skin effect.
\end{abstract}

%============================================================================
\maketitle
%============================================================================
\section{Introduction}
\label{sec:Introdution}
%============================================================================

Driven quantum many-body systems have emerged as a fruitful arena for designing nonequilibrium phases of matter, such as discrete time crystals~\cite{PhysRevLett.117.090402, PhysRevLett.116.250401}, anomalous Floquet topological phases~\cite{PhysRevX.3.031005, PhysRevX.6.021013}, and Floquet prethermal phases~\cite{PhysRevX.7.011026}.
On the experimental side, a strong motivation to study such systems stems from their ability to be realized on diverse platforms, ranging from cold atoms in optical lattices to solid state systems and photonics \cite{Wang_2013, PhysRevLett.99.220403, Rechtsman_2013}. 
However, a pivotal issue is that they generically heat up to infinite-temperature states~\cite{PhysRevX.4.041048, PhysRevE.90.012110, PhysRevX.10.011030}.
While there are exceptions to this fate---such as driven integrable models that tend to periodic Gibbs ensembles at late times~\cite{PhysRevLett.112.150401} and many-body localized systems~\cite{PhysRevLett.114.140401}---understanding nonheating late-time dynamics of driven quantum many-body systems remains a challenging task.

On the other hand, dissipative quantum systems are expected to follow an irreversible nonunitary evolution towards a steady state at late times.
This has motivated the use of tailored dissipation to prepare and control quantum many-body states~\cite{PhysRevA.59.2468, PhysRevA.78.042307, Diehl_2008}.
The competition between dissipation and drives has recently been studied for its potential ability to stabilize a variety of nonequilibrium steady states that are protected against heating and parametric instabilities~\cite{Zhu_2019, Mori_2023, PhysRevA.93.032121, PhysRevLett.114.130402, PhysRevLett.126.180503, PhysRevX.7.011016, PhysRevA.92.023815, PhysRevResearch.3.023100, 10.21468/SciPostPhys.13.5.104, PhysRevA.110.L010202}. 

\begin{figure}[t]
\centering
\includegraphics[width=0.35\textwidth]{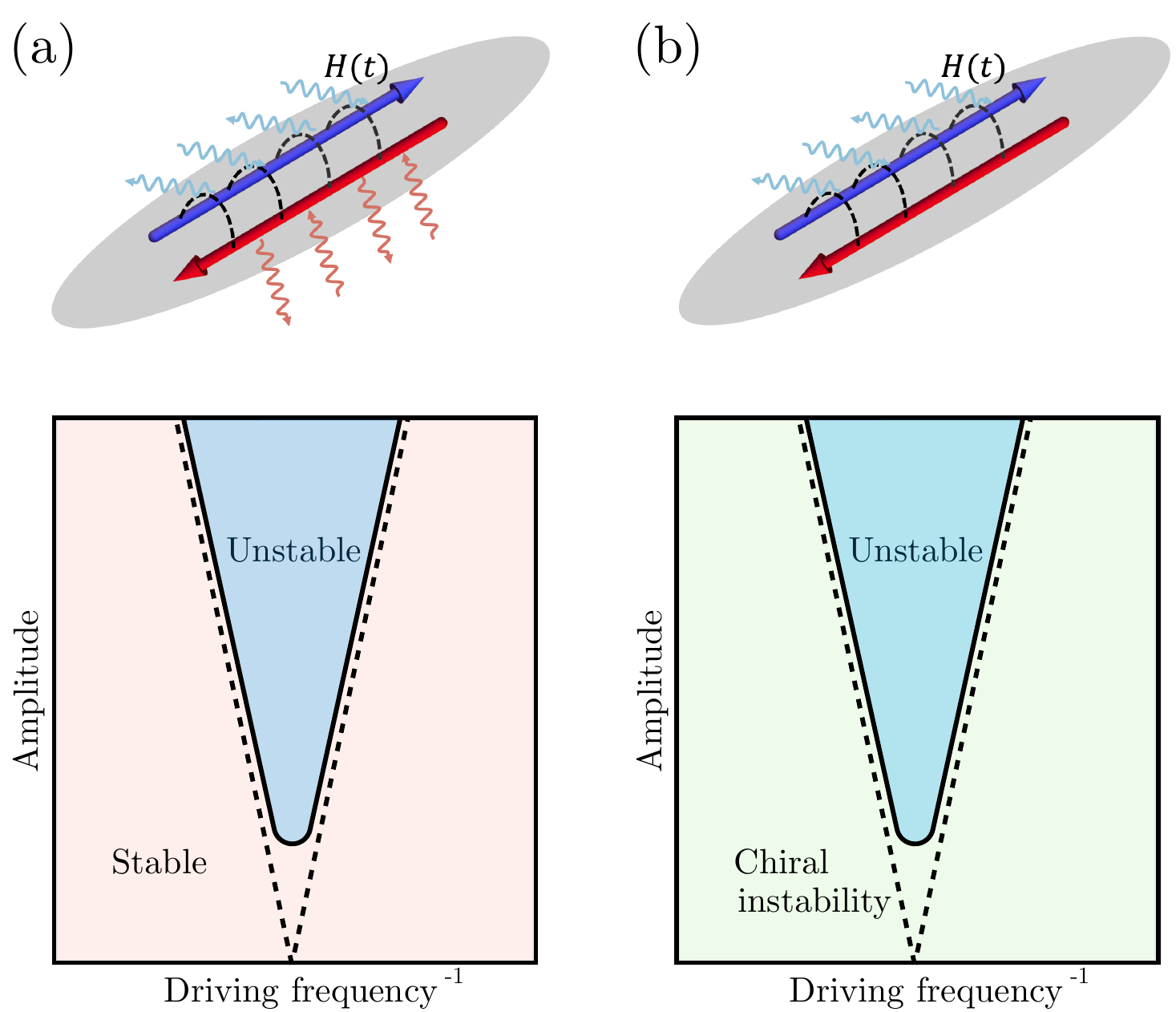}
\caption{Sketch of a driven-dissipative TLL, decomposed into right- and left-moving bosons, and evolving under a time-periodic Hamiltonian $H(t)$ which modulates the coupling between modes of opposite momenta.
The driven TLL is coupled to a thermal bath with a time-periodic dissipation $\gamma_t$ in two ways: (a) A symmetric dissipation, where both right and left movers are coupled to the bath.
In this case, the parametric resonance of the driven TLL is partially stabilized (the resonance boundary in the absence of dissipation is shown as a dashed line).
(b) A chiral dissipation, where only right movers are coupled to the bath. In this case, the driven TLL is always unstable, with a transition from a conventional \emph{symmetric parametric instability} to a new \emph{chiral parametric instability}.}
\label{fig:nonunitaryenergy}
\end{figure}

An outstanding framework to analytically investigate quantum many-body dynamics is given by Tomonaga-Luttinger-liquid (TLL) theory~\cite{Haldane:1981llt}.
This is motivated by its emergence as the effective low-energy description for a large class of gapless quantum many-body systems in 1+1 dimensions in terms of right- and left-moving bosonic excitations.
Examples include the Luttinger model~\cite{Tomonaga:1950, Luttinger:1963, MattisLieb:1965} of interacting massless fermions, the Lieb-Liniger model~\cite{LiebLiniger:1963} of interacting massless bosons, gapless spin chains~\cite{SchulzCunibertiPieri:2000, Giamarchi:2003}, and cold atomic gases~\cite{RevModPhys.83.1405}.
In particular, interaction quenches~\cite{NgoDinhEtAl:2013iq, 10.1088/1572, Cazalilla:2006iq, CazalillaIucci:2009qq, SachdevaEtAl:2014iq, KarraschEtAl:2012llu, DoraEtAl:2013bangbang_orthocata, DoraEtAl:2016entangle_Lecho, LLMM1:2017qLm, RuggieroEtAl:2021tpi, RuggieroEtAl:2021qTLL, PhysRevLett.124.136802, Moosavi:DBdG_iQLs:2023} and driven dynamics~\cite{PhysRevB.86.054304, PhysRevLett.126.243401, DLMT:2023, PhysRevB.110.155104} of TLLs have been thoroughly studied, leading to predictions for the nonequilibrium dynamics in gapless quantum many-body systems.
Moreover, the effect of dissipation has recently been investigated in TLLs to engineer tractable steady states~\cite{PhysRevLett.124.136401, PhysRevB.107.125149, PhysRevB.107.045110, PhysRevB.105.205125, PhysRevB.108.064312}.
Yet, the competition between drives and dissipation has so far not been explored in the context of TLLs.

In this work, we investigate the interplay between periodic (Floquet) drives and dissipation in TLLs, and demonstrate that it leads to new types of driven-dissipative phase transitions. 
Specifically, we design an exactly solvable dissipative drive by employing a Floquet-Lindblad approach~\cite{PhysRevA.97.062121, PhysRevA.99.022105, PhysRevB.109.184309, PhysRevB.101.100301}.
The drive consists in periodically modulating the Luttinger parameter, which would naturally lead to parametric resonances in the bosonic picture~\cite{DLMT:2023, PhysRevLett.126.243401}.
By tailoring specific types of dissipation, we demonstrate that the parametric resonances can be, on the one hand, partially stabilized, see Fig.~\ref{fig:nonunitaryenergy}(a), and on the other hand, split into different types.
In particular, we design a chiral dissipation that couples only to one species of the right/left-moving bosons, allowing for a new type of parametric instabilities where only one chirality of the quasiparticles (in the rotating time-independent frame) is amplified, see Fig.~\ref{fig:nonunitaryenergy}(b).
Modifying the strength of this chiral dissipation gives rise to a nonequilibrium phase transition between a conventional \emph{symmetric parametric instability} and a \emph{chiral parametric instability}.
In the latter, as one chirality is stabilized while the other is amplified, there is an imbalance, which bears similarities to other nonreciprocal phenomena \cite{FruchartEtAl:2021}, in particular with the non-Hermitian skin effect \cite{PhysRevLett.77.570, PhysRevLett.121.086803}. 

The rest of the paper is organized as follows.
In Sec.~\ref{sec:DDPTs_in_TLLs}, we investigate the driven-dissipative dynamics of TLLs with the dissipation coupling both right- and left-moving bosons to a thermal bath.
In Sec.~\ref{sec:CPI}, we turn to the case of a chiral dissipation, giving rise to transitions between symmetric and chiral parametric instabilities.
A summary of our work as well as an outlook for future research are given in Sec.~\ref{sec:Conclusion}.
Technical details are deferred to a number of appendices included in the Supplemental Material (SM)~\cite{SM}.

%============================================================================
\section{Driven-dissipative phase transitions in quantum liquids}
\label{sec:DDPTs_in_TLLs}
%============================================================================
\subsection{Driven-dissipative Tomonaga-Luttinger liquids}
%----------------------------------------------------------------------------

We begin by recalling some properties of TLL theory with periodic boundary conditions.
The model can be written in terms of a compactified bosonic field $\varphi(x)$ (with values modulo $2\pi$) and its conjugate field $\Pi(x)$ for $x \in [-L/2, L/2]$, where $[\partial_x \varphi(x), \Pi(y)] = i \delta'(x-y)$ and all quantities are periodic in the system size $L$.
For simplicity, we set $L = 2\pi$, meaning that $x$ plays the role of an angle coordinate and the system lies on the unit circle.
The Hamiltonian then reads (setting $\hbar = 1$)
\begin{equation}
\label{eq:H_TLL}
H = \frac{v}{2\pi} \int_{-\pi}^{\pi} \dd x\, \! \wick{ \biggl( \frac{1}{K} [\pi \Pi(x)]^2 + K [\partial_x \varphi(x)]^2 \biggr) },
\end{equation}
where $\wick{\cdots}$ indicates Wick ordering, $v > 0$ denotes the propagation velocity, and $K > 0$ is the so-called Luttinger parameter.
The latter encodes the interactions in the effectively described systems in the TLL universality class.
For instance, for the Lieb-Lininger model, $K = 1$ corresponds to the strongly interacting Tonks–Girardeau limit, while for fermions and the $\textrm{XXZ}$ spin chains, $K \neq 1$ corresponds to nonzero interactions or nonzero anisotropy, respectively. 

The importance of TLL theory in 1+1 dimensions makes it an ideal framework to study nonequilibrium phenomena in one-dimensional critical quantum many-body systems.
In this paper, we focus on Floquet-driven scenarios, described by a time-dependent Hamiltonian $H(t)$ of the form \eqref{eq:H_TLL} with $v$ and $K$ replaced by time-dependent functions $v(t)$ and $K(t)$.
Both $v(t)$ and $K(t)$ are periodically modulated, and for convenience we introduce $\epsilon_t$ and $\lambda_t$ so that
\begin{equation}
v(t) K(t) = \epsilon_t - \lambda_t,
\qquad
\frac{v(t)}{K(t)} = \epsilon_t + \lambda_t,
\end{equation}
with the condition $\epsilon_t > |\lambda_t|$.
The time-dependent TLL Hamiltonian $H(t)$ can then be decomposed as follows in terms of right (left)-moving bosonic operators $a_q\pdag = a_{-q}^\dagger$ ($\bar{a}_q\pdag = \bar{a}_{-q}^\dagger$) labeled by momenta $q = \pm 1, \pm 2, \pm 3, \ldots$ and satisfying $[a_{q}, a_{q'}] = \sgn(q) \delta_{q + q',0} = [\bar{a}_{q}, \bar{a}_{q'}]$ for $q, q' \neq 0$ \footnote{The TLL model has two conserved $\textrm{U}(1)$ currents, a vector current and an axial current, with components $\rho(x)$, $j(x)$ and $\rho_{5}(x)$, $j_{5}(x)$, respectively, and the Hamiltonian \eqref{eq:H_TLL} is symmetric under exchanging $K \leftrightarrow K^{-1}$ and $(\rho, j) \leftrightarrow (\rho_{5}, j_{5})$.
These components can be expressed in terms of the bosonic fields $\varphi(x)$, $\Pi(x)$ or alternatively using new right/left-moving bosonic fields $\rho_{\pm}(x) = \rho_{\pm}(x)^{\dagger}$.
Concretely:
$\rho(x) = \Pi(x) = \rho_{+}(x) + \rho_{-}(x)$,
$\rho_{5}(x) = - \partial_x \varphi(x)/\pi = \rho_{+}(x) - \rho_{-}(x)$,
$j(x) = v K \rho_{5}(x)$,
and
$j_{5}(x) = v K^{-1} \rho(x)$.
We define the corresponding operators $a_q$, $\bar{a}_{q}$ in momentum space as follows:
$\rho_{+}(x) = \frac{1}{L} \sum_{q \in (2\pi/L) \mathbb{Z}} e^{i qx} \Bigl( a_q \sqrt{\frac{L|q|}{2\pi}} + Q_{+} \delta_{q,0} \Bigr)$
and
$\rho_{-}(x) = \frac{1}{L} \sum_{q \in (2\pi/L) \mathbb{Z}} e^{-i qx} \Bigl( \bar{a}_q \sqrt{\frac{L|q|}{2\pi}} + Q_{-} \delta_{q,0} \Bigr)$,
where $Q_{\pm}$ are zero modes, interpreted as chiral particle number operators.
Setting $L = 2\pi$, the above transforms allow one to formally obtain \eqref{eq:Ht_aabar} (up to omitting zero modes) from \eqref{eq:H_TLL} with $v$, $K$ replaced by $v(t)$, $K(t)$.}:
\begin{equation}
\label{eq:Ht_aabar}
H(t) = \sum_{q>0} q \Bigl( \epsilon_t \bigl[ a_q^{\dagger}a_q\pdag + \bar{a}_q^{\dagger}\bar{a}_q\pdag \bigr] + \lambda_t \bigl[ a_q\pdag \bar{a}_q\pdag + a_q^{\dagger}\bar{a}_q^{\dagger} \bigr] \Bigr),
\end{equation}
up to omitted contributions from zero modes.
(The latter require a separate treatment, but as they commute with every other quantity introduced and since they will not play any role for our considerations, we only deal with driven nonzero modes and omit zero modes in what follows.)
As is manifest from \eqref{eq:Ht_aabar}, different momentum modes are uncoupled, while the Hamiltonian is only quasifree due to the time-dependent interaction $\lambda_t$ between right and left movers with the same momentum $q$.
Taking $\lambda_t = 0$ as the reference, we note that the Wick ordering in \eqref{eq:Ht_aabar} was taken with respect to the noninteracting ground state.
Last, we note that only positive momentum labels $q$ will appear from here on.

For concrete numerical calculations, we will pick
\begin{equation}
\label{eq:ot_lt}
\epsilon_t = v_0 \bigl[ 1 + B + A\cos(\omega t) \bigr],
\quad
\lambda_t = v_0 \bigl[ B + A\cos(\omega t) \bigr]
\end{equation}
for real dimensionless parameters $A$ and $B$ satisfying $1 + 2B > 2|A|$, where $v_0 > 0$ denotes the propagation velocity without interactions ($A = B = 0$) and $\omega = 2\pi / T$ is the angular frequency of the drive with period $T$.
However, our analytical results apply for general $T$-periodic $\epsilon_t$ and $\lambda_t$ as long as $\epsilon_t > |\lambda_t|$.

At any time $t$, the nonzero modes of the Hamiltonian $H(t)$ can be diagonalized by moving to a rotating frame by an instantaneous Bogoliubov transformation:
$\tilde{a}_q\pdag = \cosh(\nu_t) a_q\pdag + \sinh(\nu_t) \bar{a}_q^{\dagger}$ and $\tilde{\bar{a}}_q\pdag = \cosh(\nu_t) \bar{a}_q\pdag + \sinh(\nu_t) a_q^{\dagger}$ for $q \neq 0$ with $\nu_t$ chosen so that $\tanh(2\nu_t) = \lambda_t/\epsilon_t$.
This defines time-dependent plasmon operators, $\tilde{a}_q$ and $\tilde{\bar{a}}_q$, that satisfy the same instantaneous commutation relations as the original bosons $a_q$ and $\bar{a}_q$.

The problem of studying the dynamics of a closed periodically driven TLL can be solved using an underlying $\SU(1,1)$ symmetry of $H(t)$.
Indeed, the Floquet Hamiltonian $H_F$, which encodes the stroboscopic evolution of local observables, can be expressed as a sum over $q$ of linear combinations of $\su(1,1)$ generators built from the bosonic operators $a_q$, $\bar{a}_q$; see, e.g.,~\cite{DLMT:2023}.
For later usage, we express these $\su(1,1)$ generators as superoperators,
\begin{equation}
\label{eq:SU11_supops}
\begin{gathered}
H_{q,0} = -i[a_q^{\dagger} a_q\pdag + \bar{a}_q^{\dagger} \bar{a}_q\pdag, \bullet], \\
H_{q,1}=-2i[a_q\pdag \bar{a}_q\pdag, \bullet], \quad
H_{q,2}=-2i [a_q^{\dagger} \bar{a}_q^{\dagger}, \bullet],
\end{gathered}
\end{equation}
with $\bullet$ indicating that they act on operators in the theory, e.g., $H_{q,0} \mathcal{O} = -i[a_q^{\dagger} a_q\pdag + \bar{a}_q^{\dagger} \bar{a}_q\pdag, \mathcal{O}]$ for an operator $\mathcal{O}$.
Representing $H_F$ for each mode in terms of the generators allows one to build an invariant~\cite{DLMT:2023} that delineates two distinct dynamical phases: a stable phase, where observables oscillate in time, and an unstable phase, where right and left movers both get parametrically amplified, leading to an exponential growth of energy.
These two phases depend on the amplitude and the frequency $(A, \omega/q)$ of the drive, as discussed in \cite{PhysRevB.86.054304, PhysRevLett.126.243401, DLMT:2023} and also shown below in Fig.~\ref{fig:stability_detailedbalance}(a).
As the driving Hamiltonian $H(t)$ consists of a sum of uncoupled parametric oscillators, each mode $q$ may be independently amplified, which generically leads to an overall unstable phase at any driving frequency $\omega$ when all (infinitely many) bosonic modes are taken into account.

Our goal is to understand the fate of the aforementioned dynamical phases, as well as phase transitions between them, in the presence of a time-periodic dissipative coupling synchronized with the Floquet drive.
In particular, we aim to address whether a finite amount of dissipation may be enough to stabilize \textit{all} bosonic modes at once.
While this is in general a formidable task, we make analytical progress by tailoring a dissipation that does not couple the different momentum modes, but is engineered so the strength is proportional to the momentum $q$ for each mode, implying a nonlocal dissipation in position space.

The driven-dissipative dynamics of the system can be formulated using a time-dependent master equation
\begin{equation}
\label{eq:masterequation}
\partial_t \rho_t = \mathcal{L}_t \rho_t = -i [H(t), \rho_t] + \mathcal{D}_t,
\end{equation}
where $\rho_{t}$ denotes the density matrix describing the state of the system at time $t$ and $\mathcal{D}_t$ is the dissipative part of the Liouvillian.
The first term corresponds precisely to the Hamiltonian \eqref{eq:Ht_aabar} expressed in terms of the $\su(1,1)$ superoperators \eqref{eq:SU11_supops}.
As to the second term, the dissipative part is taken to be
\begin{multline}
\label{eq:dissipationsym}
\mathcal{D}_t
= \sum_{q>0} \gamma_t q \Bigl[ (\mathfrak{n}_{q}+1) (D_{q, 1} + \bar{D}_{q, 1}) + \mathfrak{n}_{q} (D_{q, 2} + \bar{D}_{q, 2}) \\ - \mathfrak{m}_{q} (D_{q, 3}+\bar{D}_{q, 3}) - \mathfrak{m}_{q} (D_{q, 4}+\bar{D}_{q, 4}) \Bigr]
\end{multline}
for real quantities $\mathfrak{n}_{q}$, $\mathfrak{m}_{q}$, and $\gamma_t$ together with the superoperators
{\small%
\begin{equation}
\label{eq:D_supops}
\begin{aligned}
    D_{q, 1}
    & = a_q\pdag \bullet a_q^{\dagger}
        - \frac{1}{2} \{a_q^{\dagger} a_q\pdag, \bullet\},
    & \bar{D}_{q, 1}
    & = \bar{a}_q\pdag \bullet \bar{a}_q^{\dagger}
        - \frac{1}{2} \{\bar{a}_q^{\dagger} \bar{a}_q\pdag, \bullet\}, \\
    D_{q, 2}
    & = a_q^{\dagger} \bullet a_q\pdag
        - \frac{1}{2} \{a_q\pdag a_q^{\dagger}, \bullet\},
    & \bar{D}_{q, 2}
    & = \bar{a}_q^{\dagger} \bullet \bar{a}_q\pdag
        - \frac{1}{2} \{\bar{a}_q\pdag \bar{a}_q^{\dagger}, \bullet\}, \\
    D_{q, 3}
    & = \bar{a}_q^{\dagger} \bullet a_q^{\dagger}
        - \frac{1}{2} \{a_q^{\dagger} \bar{a}_q^{\dagger}, \bullet\},
    & \bar{D}_{q, 3}
    & = a_q^{\dagger} \bullet \bar{a}_q^{\dagger}
        - \frac{1}{2} \{a_q^{\dagger} \bar{a}_q^{\dagger}, \bullet\}, \\ 
    D_{q, 4}
    & = a_q\pdag \bullet \bar{a}_q\pdag
        - \frac{1}{2} \{a_q\pdag \bar{a}_q\pdag, \bullet\},
    & \bar{D}_{q, 4}
    & = \bar{a}_q\pdag \bullet a_q\pdag
        - \frac{1}{2} \{a_q\pdag \bar{a}_q\pdag, \bullet\},
\end{aligned}%
\end{equation}
}%
which are quadratic in the bosons.
Here, the overall time-periodic dissipation strength is $\gamma_t q$, where $\gamma_t$ is assumed to be a nonnegative function with the same periodicity $T = 2\pi/\omega$ as for $\lambda_t$ and $\epsilon_t$.
For concrete numerical calculations, we will use
\begin{equation}
\label{eq:gammatdissp}
\gamma_t = \gamma [1 + \cos(\omega t)],
\end{equation}
with a nonnegative parameter $\gamma$, but our discussion applies to any $T$-periodic dissipation strength $\gamma_t q \geq 0$.

The master equation~\eqref{eq:masterequation} can be seen as a Lindblad equation~\cite{Lindblad:1976} by taking the set of jump operators to be $\{ \tilde{a}_q\pdag, \tilde{\bar{a}}_q\pdag, \tilde{a}_q^{\dagger}, \tilde{\bar{a}}_q^{\dagger} \}_{q>0}$.
Importantly, in this section, we will pick the coefficients of the Liouvillian $\mathcal{L}_t$ such that it satisfies detailed balance.
Concretely, we choose
\begin{equation}
\begin{aligned}
\label{eq:boseinsteinsqueezed}
\mathfrak{n}_{q}
& = \frac{\epsilon_t}{2\Omega_t} \coth\left( \frac{\beta q \Omega_t}{2} \right) - \frac{1}{2}, \\
\mathfrak{m}_{q}
& = - \frac{\lambda_t}{2 \Omega_t} \coth\left( \frac{\beta q \Omega_t}{2} \right),
\end{aligned}
\end{equation}
where $\Omega_t = \sqrt{\epsilon_t^2 - |\lambda_t|^2}$ and $\beta$ is the inverse temperature of the bath.

To solve the time-dependent Lindblad equation \eqref{eq:masterequation}, we construct a rotating frame $\tilde{\rho}_t = W(t) \rho_t$
implemented by a time-periodic superoperator $W(t)$ such that the effective Liouvillian
\begin{equation}
\label{eq:cL_tilde}
\tilde{\mathcal{L}}
= W(t)\mathcal{L}_t W(t)^{-1} + \frac{dW(t)}{dt} W(t)^{-1}
\end{equation}
becomes time-independent.
The stroboscopic time evolution of the initial density matrix $\rho_{t_0}$ at some time $t_0$ is then encoded in the Floquet Liouvillian
\begin{equation}
\label{eq:cL_F}
\mathcal{L}_F (t) = W(t)^{-1} \tilde{\mathcal{L}} W(t)
\end{equation}
evaluated at $t = t_0$.
Indeed, the latter allows one to write
\begin{equation}
\label{eq:rotatingevolution}
\rho_{t_0+ nT}
= e^{nT\mathcal{L}_F(t_0)} \rho_{t_0},
\end{equation}
which is precisely the stroboscopic time evolution.

While an explicit construction of the rotating-frame superoperator $W(t)$ is generically out of reach, our problem decomposes into individual modes, such that $W(t)$ may be explicitly constructed as a product over $q$ of commuting superoperators.
To do so, we follow the algebraic approach in~\cite{PhysRevA.97.062121, PhysRevA.99.022105} for the driven-dissipative quantum harmonic oscillator.
The key observation is that our Liouvillian $\mathcal{L}_t$ is the sum of superoperators that form a \textit{closed algebra}.
In fact, the unitary part of the Liouvillian lies in an $\su(1,1)$ algebra, which together with the pairs of superoperators in the dissipative part $\mathcal{D}_t$ constitutes a 7-dimensional algebra.
Crucially, this division into unitary and dissipative parts is precisely the decomposition of the full algebra into a semisimple subalgebra, spanned by $\{ H_{q,0}, H_{q,1}, H_{q,2} \}$, and a solvable ideal, spanned by $\{ D_{q, 1}+\bar{D}_{q, 1}, D_{q, 2} + \bar{D}_{q, 2}, D_{q, 3}+\bar{D}_{q, 3}, D_{q, 4}+\bar{D}_{q, 4} \}$.
This allows for a decomposition of the rotating-frame superoperator into $W(t) = \prod_{q>0} V_q(t) U_q(t)$ with
\begin{subequations}
\label{UqVq}
\begin{align}
U_q(t)
& = \prod_{i=1}^3 e^{f_{q, i} H_{q, i}} ,\\
V_q(t)
& = \prod_{i=1}^4 e^{g_{q, i} (D_{q, i}+\bar{D}_{q, i})},
\end{align}
\end{subequations}
where the coefficients $f_{q, i} = f_{q, i}(t)$ and $g_{q, i} = g_{q, i}(t)$ are dimensionless time-periodic functions chosen so that $\tilde{\mathcal{L}}$ is time-independent through \eqref{eq:cL_tilde}. 
The procedure to fix these coefficients follows~\cite{PhysRevA.97.062121}; see Appendix~\ref{app:rotating_frame} in the SM~\cite{SM} for details.
The effective Liouvillian can eventually be expressed in terms of the driving parameters and the dissipation strength as
\begin{equation}
\label{eq:effective_Liouvillian}
\tilde{\mathcal{L}}
= \sum_{q>0} q
  \Bigl[ \bar{\Lambda}_q H_{q, 0} + \bar{\gamma} (D_{q, 1}+\bar{D}_{q, 1}) \Bigr],
\end{equation}
where $\bar{\gamma}$ and $\bar{\Lambda}_q$ are time-averaged quantities over a full period $T = 2\pi/\omega$ that encode the effects of the dissipation and the Floquet drive, respectively: they can be expressed as
\begin{equation}
\label{eq:gammaBar}
\bar{\gamma}
= \frac{1}{T} \int_0^T \dd t\, \gamma_t
\end{equation}
and
\begin{equation}
\label{eq:Lambdaq}
\bar{\Lambda}_q
= \frac{1}{T}\int_0^T \dd t\, \bigl( \epsilon_t + 2i \lambda_t f_{q, 2} \bigr),
\end{equation}
with $f_{q,2} = f_{q,2}(t)$ a time-periodic solution of the Riccati equation
\begin{equation}
\dot{f}_{q, 2}
= - \frac{q\lambda_t}{2} - 2i q\epsilon_t f_{q, 2} + 2 q \lambda_t f_{q, 2}^2.
\label{riccatieq:fort2}
\end{equation}
As we shall see, the solution of \eqref{riccatieq:fort2} can lead to parametric instabilities.
In particular, one may note that these instabilities originate solely from the driven coupling $\lambda_t$ between right and left movers.

%----------------------------------------------------------------------------
\subsection{Phase diagram and parametric instability}
\label{subsec:SPI:pd_pi}
%----------------------------------------------------------------------------

While the stroboscopic time evolution is encoded in the Floquet Liouvillian \eqref{eq:cL_F}, its spectrum coincides with the time-independent effective Liouvillian \eqref{eq:cL_tilde} since they are related through a similarity transformation.
It is therefore instructive to study the spectrum of~\eqref{eq:effective_Liouvillian} in order to analyze the stability of the TLL and understand how the stable-to-unstable phase transition of the driven TLL is altered by dissipation.

The spectrum of $\tilde{\mathcal{L}}$ can be found analytically and amounts to two uncoupled harmonic oscillators for each mode $q$, one for each chirality (see Appendix~\ref{app:HOspectrum} in the SM~\cite{SM}):
\begin{equation}
\label{symmetric_spectrum}
\begin{aligned}
E_{n_1, n_2, m_1, m_2}^{(q)}
& = i q \bar{\Lambda}_q (m_1 - n_1 + m_2 - n_2) \\
& \quad - \frac{q\bar{\gamma}}{2} (m_1 + n_1 + m_2 + n_2)
\end{aligned}
\end{equation}
for nonnegative integers $n_{1}, m_{1}$ corresponding to right movers and $n_{2}, m_{2}$ to left movers.

\begin{figure}[t]
\centering
\includegraphics[width=\linewidth]{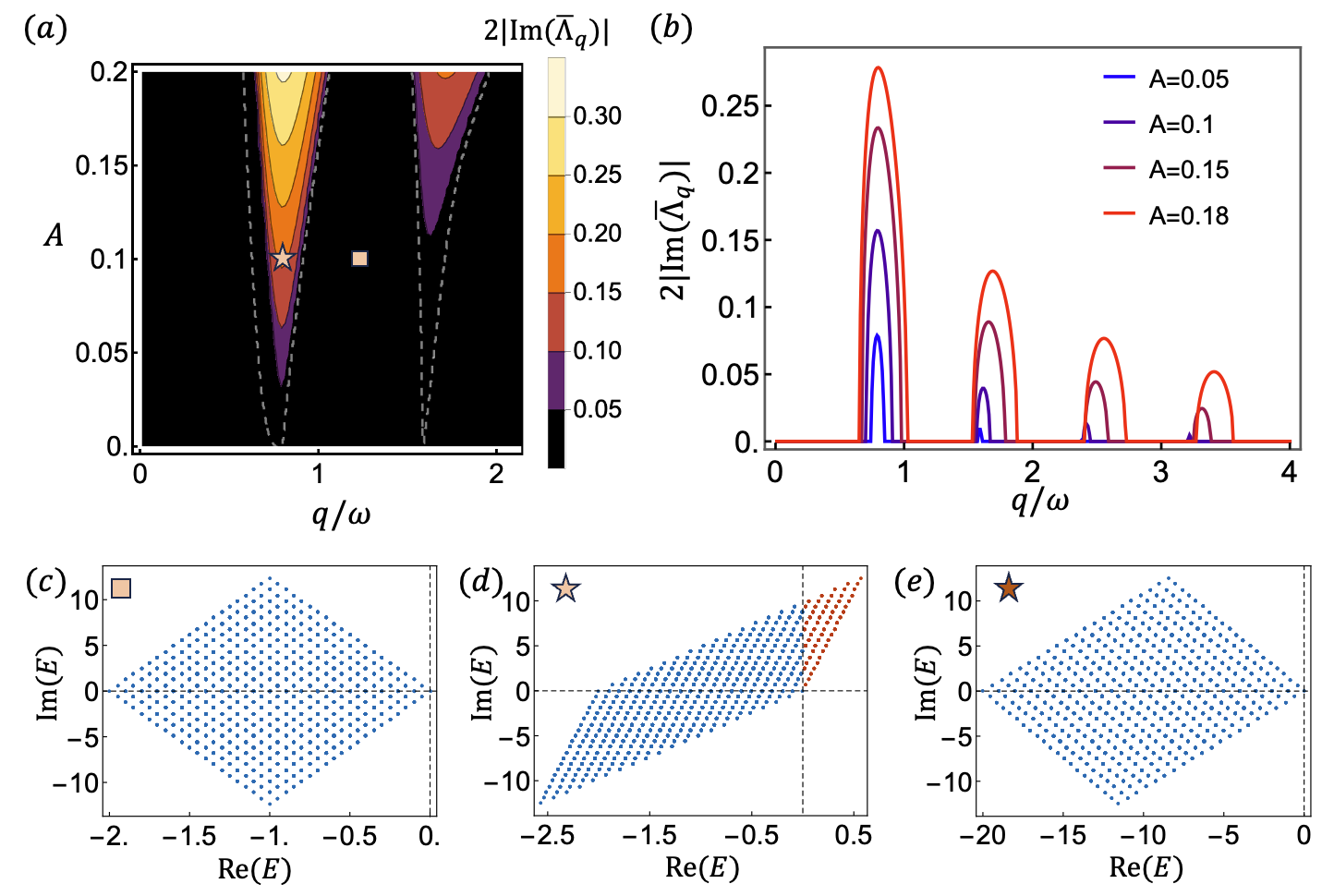}
\caption{Properties of the driven-dissipative phase transition between stable and unstable phases.
(a) Plot of $2|\Im(\bar{\Lambda}_q)|$ as a function of the driving parameters $(A, \omega/q)$ for
a single mode $q$ of a TLL driven according to \eqref{eq:ot_lt} with $B = -0.3$, $v_0 = 1$.
Different contours of $2|\Im(\bar{\Lambda}_q)|$ correspond to unstable regions for different values of the dissipation $\gamma$ in \eqref{eq:gammatdissp} through the condition \eqref{eq:symmetric_stab}.
The phase boundary in the absence of dissipation ($\gamma = 0$) is shown as a dashed white curve.
(b) Without dissipation, transitions between phases can be seen as nonanalyticities in $2|\Im(\bar{\Lambda}_q)|$ viewed as a function of $q/\omega$, plotted along rays in (a) with different values of $A$.
With dissipation, the nonanalyticities are instead those when the $2|\Im(\bar{\Lambda}_q)|$ curve intersects the horizontal line $\gamma > 0$.
(c-e) The Liouvillian spectrum \eqref{symmetric_spectrum} at different points (square, star) in (a) for $\gamma = 0.1$ (c-d) and $\gamma = 1$ (e).
(c) The spectrum has a purely nonpositive real part (indicated in blue) in the stable phase.
(d) It develops a positive real part (indicated in red) in the unstable phase.
(e) This positive part is suppressed for sufficiently large values of $\gamma$, stabilizing the TLL.}
\label{fig:stability_detailedbalance}
\end{figure}

When adding a nonzero time-averaged dissipation, $\bar{\gamma} \neq 0$, the real part of the Liouvillian spectrum is always nonpositive as long as the stability condition 
\begin{equation}
\label{eq:symmetric_stab}
\bar{\gamma} \geq 2|\Im (\bar{\Lambda}_q)|
\end{equation}
is satisfied.
If the inequality is violated, then the system becomes unstable as the dissipation is not strong enough to avoid a given mode from being exponentially amplified.
This can be used to infer the phase diagram by computing the right-hand side of \eqref{eq:symmetric_stab} for different driving parameters $(A, \omega/q)$, as shown in Fig.~\ref{fig:stability_detailedbalance}(a), and comparing the value at each point with a given $\bar{\gamma}$.
We stress that, while trace-preserving Liouvillians in finite-dimensional Hilbert spaces necessarily have a negative real part of their spectrum, such a constraint is not required for infinite-dimensional Hilbert spaces, which can lead to such instabilities in bosonic systems~\cite{Thompson_2023}.
This is analogous to parametric instabilities in the nondissipative limit, which occur whenever the spectrum of the Floquet Hamiltonian is not bounded from below.

It is important to note that the stability condition \eqref{eq:symmetric_stab} is ``scale invariant'' under changes in $q$ as long as $\omega$ is rescaled by the same factor.
This can be readily shown by examining \eqref{eq:Lambdaq}: the only dependence on $q$ in $\bar{\Lambda}_q$ comes from the solution $f_{q, 2}$ to \eqref{riccatieq:fort2}.
By changing variable from $t$ to $\phi = \omega t$, the Riccati equation becomes
\begin{equation}
\label{eq:riccati_phi}
\frac{\dd f_{q, 2}}{\dd \phi} + \frac{q}{\omega} \left( \frac{\lambda_\phi}{2} + 2i \epsilon_\phi f_{q, 2} - 2 \lambda_\phi f_{q, 2}^2 \right) = 0,
\end{equation}
where (for ease of notation) $f_{q, 2}$, $\lambda_\phi$, and $\epsilon_\phi$ now denote $2\pi$-periodic functions of $\phi$.
It follows that $\bar{\Lambda}_q$ only depends on the ratio $\omega/q$.
This claim is not a priori obvious: the original Liouvillian $\mathcal{L}_t$ does not have such invariance, and it is only in the rotating frame that the resulting stability condition develops this ``scale invariance.''

The stability condition \eqref{eq:symmetric_stab} was only stated for a single mode $q$.
This can be generalized to
\begin{equation}
\label{eq:symmetric_stab_all_q}
\bar{\gamma} \ge \max_{q = 1, 2, \ldots} 2| \Im(\bar{\Lambda}_{q}) |,
\end{equation}
which takes all modes into account.
By the ``scale invariance'' deduced from \eqref{eq:riccati_phi}, the stability condition \eqref{eq:symmetric_stab_all_q} for all modes can be converted into a condition for a single mode at different frequencies $\omega/q$.
This has a crucial consequence:
as illustrated in Fig.~\ref{fig:stability_detailedbalance}(b), $|\Im(\bar{\Lambda}_{q})|$ as a function of $q/\omega$ reaches a global maximum in the first instability lobe for any given driving amplitude $A$ in \eqref{eq:ot_lt}, i.e., any exponential growth of excitations is largest in the first mode. 
This implies that if the time-averaged coupling to the bath $\bar{\gamma}$ is large enough to stabilize the first resonance peak in Fig.~\ref{fig:stability_detailedbalance}(b) at a given amplitude, it will stabilize all subsequent resonance peaks.
We thus conclude that a finite amount of dissipation can stabilize the many-body parametric resonance in a driven TLL.

%----------------------------------------------------------------------------
\subsection{Time evolution of observables}
\label{subsec:te_obs}
%----------------------------------------------------------------------------

To better understand the stable-to-unstable phase transition discussed in the previous section, we show how to obtain the stroboscopic evolution of physical observables. 

From \eqref{eq:rotatingevolution}, the stroboscopic time evolution is governed by the Floquet Liouvillian \eqref{eq:cL_F} at the initial time $t_0$, which can be expressed as
\begin{align}
& \mathcal{L}_F(t_0)
= \sum_{q>0} q \biggl[ \omega^F_{q, t_0} H_{q, 0} + \frac{\lambda^F_{q, t_0}}{2} H_{q, 1} + \frac{\lambda'^F_{q, t_0}}{2} H_{q, 2} \label{cL_F_t0} \\
& \qquad + \bar{\gamma} \Bigl( \bigl[\mathfrak{n}^F_{q, t_0}+1\bigr] (D_{q, 1} + \bar{D}_{q, 1}) + \mathfrak{n}^F_{q, t_0} (D_{q, 2} + \bar{D}_{q, 2}) \nonumber \\
& \qquad - \mathfrak{m}^F_{q, t_0} (D_{q, 3} + \bar{D}_{q, 3}) - \mathfrak{m}'^F_{q, t_0} (D_{q, 4} + \bar{D}_{q, 4}) \Bigr) \biggr], \nonumber
\end{align}
with explicit formulas for the quantities $\omega^F_{q, t}$, $\lambda^F_{q, t}$, $\lambda'^F_{q, t}$, $\mathfrak{n}^F_{q, t}$, $\mathfrak{m}^F_{q, t}$, and $\mathfrak{m}'^F_{q, t}$ given in Appendix~\ref{sec:floquetliouvil} in the SM~\cite{SM}. 
We note that, in the case of detailed balance, the Floquet Liouvillian for each mode $q$ takes a similar form to that of the driven-dissipative quantum harmonic oscillator~\cite{PhysRevA.97.062121}.

\begin{figure}[t]
\centering
\includegraphics[width=\linewidth]{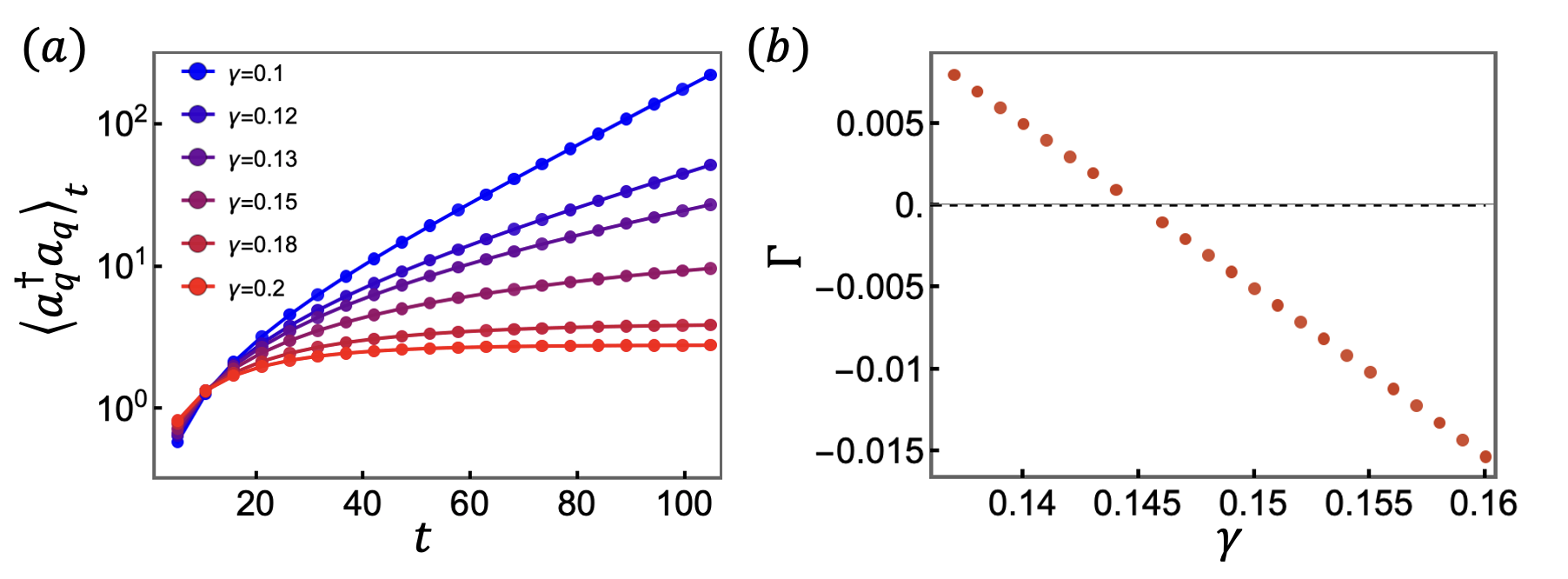}
\caption{The stroboscopic evolution of the correlator $\langle a_q^\dagger a_q\pdag \rangle_{t}$ under our drive with $\epsilon_t$ and $\lambda_t$ in \eqref{eq:ot_lt} for the case of a symmetric dissipation with $\gamma_t$ in \eqref{eq:gammatdissp}.
(a) The evolution of $\langle a_q^\dagger a_q\pdag \rangle_{t}$ for a single mode $q = 1$, obtained by numerically solving its equation of motion and plotted on a logarithmic scale for different values of the dissipation $\gamma$.
(The other parameters are $A = 0.1$, $\omega = 1.25$, $B = -0.3$, $v_0 = 1$, $\beta = 1$, and $t_0 = 0$.)
Starting from an unstable state in the nondissipative limit, the correlator exhibits exponential growth.
As $\gamma$ increases, the instability is suppressed, eventually allowing the correlator to converge to a steady-state value.
(b) The corresponding fit based on \eqref{eq:correlator_rate} showing the dependence of the rate $\Gamma$ as a function of $\gamma$, exhibiting a smooth crossover near the phase transition.}
\label{fig: correlator}
\end{figure}

Using $\mathcal{L}_F(t_0)$ in \eqref{cL_F_t0}, the equations of motion for the correlators
$\langle a_q^\dagger a_q\pdag \rangle_{t}$, $\langle \bar{a}_q^\dagger \bar{a}_q\pdag \rangle_{t}$, $\langle a_q\pdag \bar{a}_q\pdag \rangle_{t}$, and $\langle a_q^\dagger \bar{a}_q^\dagger \rangle_{t}$ 
can be analytically derived; see Appendix~\ref{sec:EOMoperators} in the SM~\cite{SM}.
Here, $\langle \mathcal{O} \rangle_{t} = \Tr \bigl[ \tilde{\rho}_{t} \mathcal{O} \bigr]$ denotes the expectation value of an operator $\mathcal{O}$ with respect to the ``dressed'' density matrix $\tilde{\rho}_{t} = e^{(t-t_0)\mathcal{L}_F(t_0)}\rho_{t_0}$, which agrees with the expectation value with respect to the ``bare'' density matrix $\rho_t$ at stroboscopic times $t = t_0 + nT$ ($n \in \mathbb{Z}$). 
To investigate the dynamics, we set the initial state as the TLL ground state and numerically solve the equation of motion for the correlator $\langle a_q^\dagger a_q\pdag \rangle_{t}$ for drives $\epsilon_t$ and $\lambda_t$ chosen as in \eqref{eq:ot_lt} and a symmetric dissipation with $\gamma_t$ as in \eqref{eq:gammatdissp}.
The results, shown in Fig.~\ref{fig: correlator}, reveal distinct dynamics for $\langle a_q^\dagger a_q\pdag \rangle_{t}$ in the stable and unstable phases.
Specifically, in the unstable regime, the correlator grows exponentially, aligning with the classification determined by the stability condition.
In contrast, within the stable regime, the correlator eventually saturates to a steady state, indicating the predominance of dissipation over parametric amplification.

The asymptotic growth (decay) rate of the correlator evolution reveals the extent of the instability (stability) of the driven TLL.
To characterize this amplifying (decaying) behavior, we fit the correlator time evolution as 
\begin{equation}
\label{eq:correlator_rate}
\langle a_q^\dagger a_q\pdag \rangle_{t}
\sim C(1-e^{\Gamma t}),
\end{equation}
with a rate $\Gamma$ and some constant $C$.
Thereby, we can extract the change of $\Gamma$ across the phase boundary, as shown in Fig.~\ref{fig: correlator}.
In the unstable phase, the rate is by definition positive, corresponding to exponential growth.
As the system approaches the phase boundary, $\Gamma$ exhibits a smooth crossover through zero, marking a continuous transition into the stable phase. 

%============================================================================
\section{Chiral parametric instabilities}
\label{sec:CPI}
%============================================================================
\subsection{Chiral dissipation}
%----------------------------------------------------------------------------

We now introduce another form of coupling to the thermal bath that goes beyond the detailed balance scenario studied in the previous section.
In particular, we aim to design a driven-dissipative system which showcases phase transitions that cannot exist in purely driven or purely dissipative settings.
To this end, we introduce a chiral coupling to the bath, modeled by the Liouvillian $\mathcal{L}_t$ in \eqref{eq:masterequation} with
\begin{equation}
\label{chiral_Liouvillian}
\mathcal{D}_t
= \sum_{q>0} q\gamma_t \Bigl[ (\mathfrak{n}_{q}+1) D_{q, 1} + \mathfrak{n}_{q} D_{q, 2} \Bigr]
\end{equation}
for $\mathfrak{n}_{q}$ given by \eqref{eq:boseinsteinsqueezed} and a nonnegative $T$-periodic function $\gamma_t$, similar to before.
This case corresponds to taking the set of jump operators to be $\{ a_q\pdag, a_q^{\dagger} \}_{q>0}$.
In particular, the main difference of~\eqref{chiral_Liouvillian} with respect to~\eqref{eq:dissipationsym} is that the Liouvillian only couples the right-moving bosons to the thermal bath.
While right and left movers do not couple in the absence of an external Floquet drive, the temporal modulation of the Luttinger parameter explicitly introduces a time-dependent coupling between the two chiralities.
Therefore, while our model does not explicitly couple the left-moving bosons to the bath, we expect that the interplay between the dissipation and the drive will effectively dissipate left-moving modes.

To study the (in)stability of such a driven-dissipative TLL, we will use a similar strategy as in Sec.~\ref{sec:DDPTs_in_TLLs}.
In fact, the superoperators appearing in~\eqref{chiral_Liouvillian} are part of a larger 11-dimensional algebra $\mathfrak{g}$ of superoperators, whose details are given in Appendix~\ref{sec:chira_dissp} in the SM~\cite{SM}.
Following the previous procedure, we decompose $\mathfrak{g}$ into an ideal and the direct sum of two $\su(1, 1)$ algebras.
It follows that, in each momentum sector $q$, the corresponding rotating-frame superoperator $W(t)$ can be expressed as a product of exponentials of the generators of these three subalgebras.

After a lengthy calculation (see Appendix~\ref{sec:chira_dissp} in the SM~\cite{SM}), the effective Liouvillian is found to take a simple form, reminiscent of \eqref{eq:effective_Liouvillian}:
\begin{equation}
\label{eq:timeindepliouvchiral}
\begin{aligned}
\tilde{\mathcal{L}} = \sum_{q>0}  q \biggl[ & \frac{1}{2} \bigl( \bar{\Lambda}_{q}^{+} + \bar{\Lambda}_{q}^{-} \bigr) H_{q, 0} + \biggl( \frac{\bar{\gamma}}{2}+i \bigl( \bar{\Lambda}_{q}^{+} - \bar{\Lambda}_{q}^{-} \bigr) \biggr) D_{q, 1} \\
& + \biggl( \frac{\bar{\gamma}}{2}-i \bigl( \bar{\Lambda}_{q}^{+} - \bar{\Lambda}_{q}^{-} \bigr) \biggr) \bar{D}_{q, 1} \biggr],
\end{aligned}
\end{equation}
where $\bar{\gamma}$ is obtained from the chiral dissipation $\gamma_t$ in the same way as in \eqref{eq:gammaBar} and the quantities $\bar{\Lambda}_{q}^{\pm}$ are defined as
\begin{equation}
\label{eq:Lambda_q_pm}
\begin{aligned}
\bar{\Lambda}_{q}^{+}
& = \frac{1}{T}\int_0^T \dd t\, \left( \epsilon_t + \frac{\gamma_t}{4i} + 2i \lambda_t f_{q,2}^{+} \right), \\
\bar{\Lambda}_{q}^{-}
& = \frac{1}{T}\int_0^T \dd t\, \left( \epsilon_t -\frac{\gamma_t}{4i} + 2i \lambda_t f_{q, 2}^{-} \right),
\end{aligned}
\end{equation}
with $f_{q,2}^{+} = f_{q,2}^{+}(t)$ and $f_{q, 2}^{-} = f_{q, 2}^{-}(t)$
time-periodic solutions of the Riccati equations
\begin{equation}
\label{eq:riccatichiral}
\begin{aligned}
\dot{f}_{q, 2}^{+}
& = - q\frac{\lambda_t}{2}
    - 2i q \left( \epsilon_t+\frac{\gamma_t}{4i} \right) f_{q,2}^{+}  + 2 q \lambda_t (f_{q,2}^{+})^2, \\
\dot{f}_{q, 2}^{-}
& = - q\frac{\lambda_t}{2}
    - 2iq \left( \epsilon_t-\frac{\gamma_t}{4i} \right) f_{q, 2}^{-} + 2q\lambda_t (f_{q, 2}^{-})^2.
\end{aligned}
\end{equation}
This effective Liouvillian only involves the superoperators $H_{q, 0}$, $D_{q, 1}$ and $\bar{D}_{q, 1}$, similarly to~\eqref{eq:effective_Liouvillian}. However, the chiral imbalance of the time-dependent Liouvillian manifests itself through unequal coefficients for $D_{q, 1}$ and $\bar{D}_{q, 1}$.
Importantly, although the original Liouvillian does not contain any term that couples the left-moving modes to the thermal bath, $\tilde{\mathcal{L}}$ explicitly contains such a term.
This is expected on physical grounds, as the Floquet drive couples the chiralities, and thus converts right movers into left movers, which are then dissipated.
In particular, removing the coupling term  (i.e., setting $\lambda_t = 0$), directly implies a vanishing of the term proportional to $\bar{D}_{q, 1}$, as expected.
In the following, we will study the physical consequences of this imbalance between different chiralities, and how it changes the nature of the driven-dissipative phase transition.

%----------------------------------------------------------------------------
\subsection{Chiral instability}
%----------------------------------------------------------------------------

The spectrum of the effective Liouvillian \eqref{eq:timeindepliouvchiral} may be obtained using a vectorization procedure similar to the one presented in Appendix~\ref{app:HOspectrum} in the SM~\cite{SM}.
It follows that
\begin{align}
E_{n_1, n_2, m_1, m_2}^{(q)}
& = \frac{i q}{2}
    \bigl( \bar{\Lambda}_{q}^{+} + \bar{\Lambda}_{q}^{-} \bigr) (m_1 - n_1 + m_2 - n_2) \nonumber \\
& \quad - \frac{i q}{2}
    \bigl( \bar{\Lambda}_{q}^{+} - \bar{\Lambda}_{q}^{-} \bigr) (m_1 + n_1 - m_2 - n_2) \nonumber \\
& \quad - \frac{q\bar{\gamma}}{4} (m_1 + n_1 + m_2 + n_2), \label{eq:spectrumchiral}
\end{align}
again for nonnegative integers $n_{1}, m_{1}$ corresponding to right movers and $n_{2}, m_{2}$ to left movers.
As in the previous section, stability for a given mode $q$ is achieved by requiring a nonpositive real part of the spectrum \eqref{eq:spectrumchiral}, leading to the condition
\begin{equation}
\label{eq:stabcondchiral}
\bar{\gamma} \geq 4 \max \bigl\{ |\Im(\bar{\Lambda}_{q}^{+})|, |\Im(\bar{\Lambda}_{q}^{-})| \bigr\}.
\end{equation}
(This can be extended to a condition for the whole system by also maximizing over $q$.)
In the case of detailed balance discussed previously, the dissipation $\gamma_t$ did not enter into the expression for $\bar{\Lambda}_q$, which implied the existence of a critical dissipation above which the system is stabilized.
However, the stability condition \eqref{eq:stabcondchiral} has a highly nonlinear dependence on the dissipation, because the $\bar{\Lambda}_{q}^{\pm}$ explicitly depend on $\gamma_t$ through the Riccati equations~\eqref{eq:riccatichiral}.
Therefore, we may expect that such a nontrivial dependence of the stability condition on $\gamma_t$ may lead to a richer set of dynamical phases as compared to the case of a symmetric bath coupling.

\begin{figure}[t]
\centering
\includegraphics[width=\linewidth]{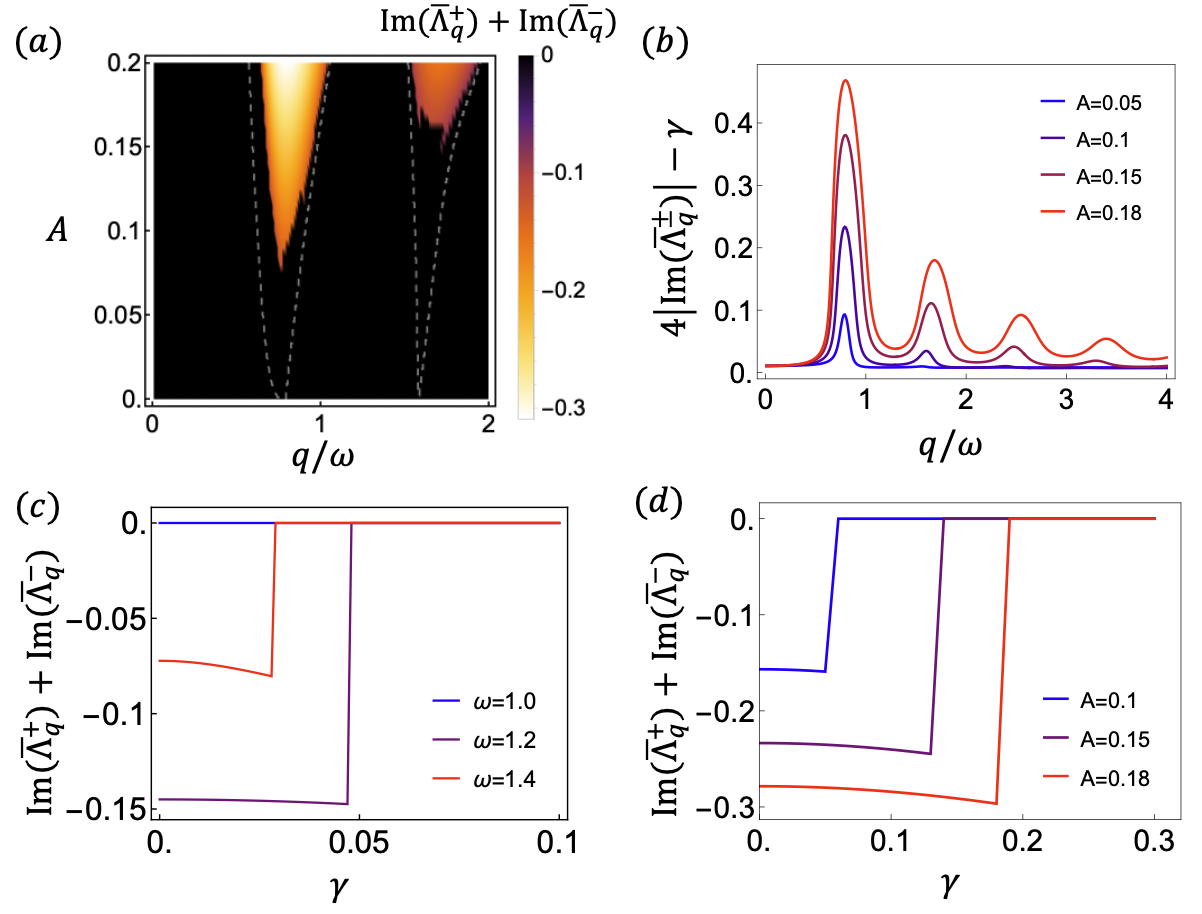}
\caption{Properties of the driven-dissipative phase transition between chiral and symmetric parametric instabilities.
(a) Plot of $\Im(\bar{\Lambda}_{q}^{+}) + \Im(\bar{\Lambda}_{q}^{-})$ as a function of the driving parameters $(A, \omega/q)$ for a single mode $q$ of a TLL with drives \eqref{eq:ot_lt} and chiral dissipation \eqref{eq:gammatdissp} for $B = -0.3$, $v_0 = 1$, and $\gamma = 0.05$.
The plot can be interpreted as a phase diagram with chiral instabilities (black regions) and symmetric instabilities (colored).
The phase boundary in the absence of dissipation ($\gamma = 0$) is shown as a dashed white curve.
(b) Plot of $4|\Im(\bar{\Lambda}_{q}^{\pm})|-\gamma$ as a function of $q/\omega$ for different values of $A$
when $\gamma = 0.1$ [all other choices as in (a)].
The curves are always positive, and the outcome is qualitatively the same for other values of the chiral dissipation $\gamma$, indicating that the stability condition \eqref{eq:stabcondchiral} is never satisfied.
(c-d) Phase transitions are instead between different instabilities, depending on whether \eqref{eq:chiralcondition1} or \eqref{eq:chiralcondition2} is satisfied, detectable as nonanalyticities in $\Im(\bar{\Lambda}_{q}^{+}) + \Im(\bar{\Lambda}_{q}^{-})$ as a function of $\gamma$, plotted for a single mode $q = 1$ in (c) when $A = 0.1$ and (d) when $\omega = 1.25$ [all other choices as in (a)].}
\label{fig:chiral_phase}
\end{figure}

\begin{figure}[t]
\centering
\includegraphics[width=\linewidth]{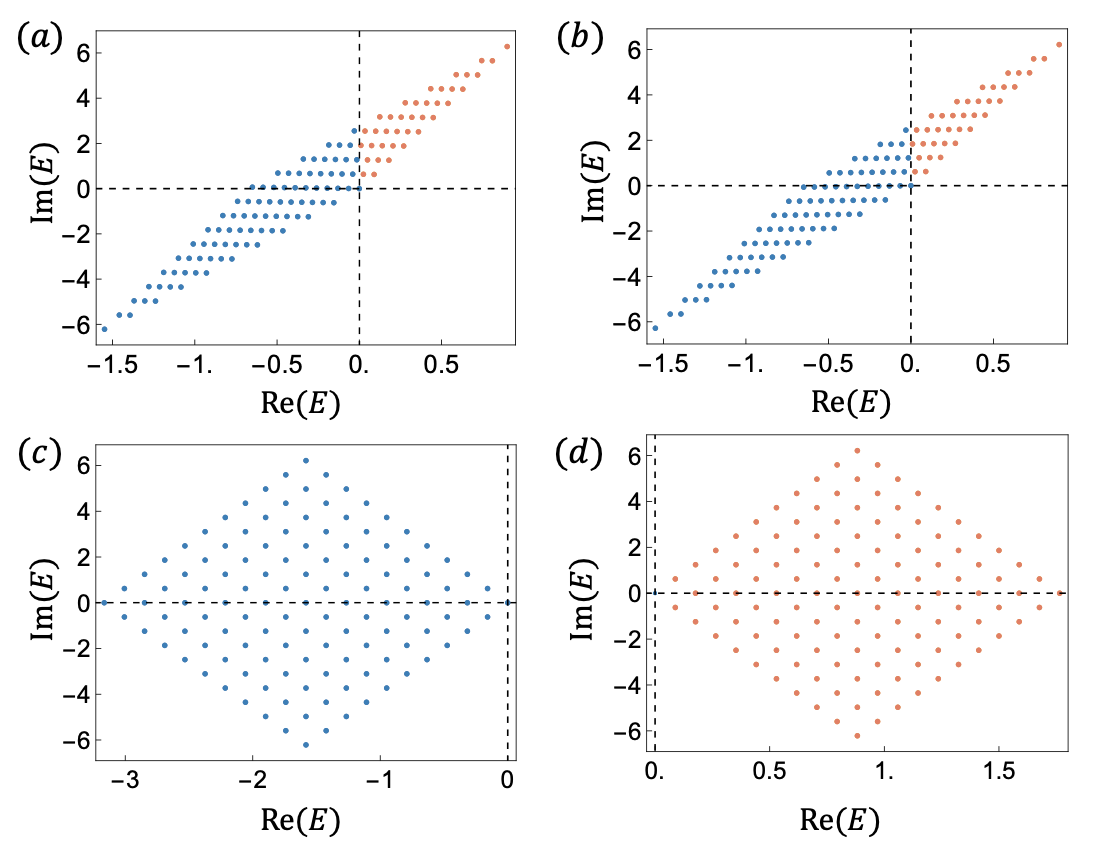}
\caption{The Liouvillian spectrum \eqref{eq:spectrumchiral} in the case of the chiral dissipation for right and left movers, plotted separately in (a) and (b) in the symmetric unstable phase and in (c) and (d) in the chiral unstable phase.
The functions $\epsilon_t$ and $\lambda_t$ are chosen as in \eqref{eq:ot_lt} and $\gamma_t$ as in \eqref{eq:gammatdissp} with $A = 0.15$, $B = -0.3$, $v_0 = 1$, $\omega = 1.25$, (a-b) $\gamma = 0.13$, and (c-d) $\gamma = 0.14$.
Only the points with $n_{1,2}$, $m_{1,2}$ ranging from 0 to 10 are included.}
\label{fig:chiral_spectrum}
\end{figure}

To investigate the emergence of new phases, we numerically solve the governing equations \eqref{eq:Lambda_q_pm} and \eqref{eq:riccatichiral} for a single momentum mode $q$ with chiral dissipation $\gamma_t$ chosen as in \eqref{eq:gammatdissp} and the drives $\epsilon_t$ and $\lambda_t$ as in \eqref{eq:ot_lt}.
These choices are assumed for the remainder of this section and used for the results plotted in Figs.~\ref{fig:chiral_phase} and~\ref{fig:chiral_spectrum}.

When the driving parameters $(A, \omega/q)$ are chosen to be in the stable phase in the absence of dissipation (cf.\ Sec.~\ref{subsec:SPI:pd_pi}), any added chiral dissipation $\gamma_t$ plays a counterintuitive role as it leads to a parametric instability whenever $\bar{\gamma} \neq 0$.
Indeed, from our numerical calculations, not only do we observe that $|\Im(\bar{\Lambda}_{q}^{+})|$ and $|\Im(\bar{\Lambda}_{q}^{-})|$ are equal, but crucially that the stability condition~\eqref{eq:stabcondchiral} is never satisfied, as can be observed from the results plotted in Fig.~\ref{fig:chiral_phase}(b).
Thus, the role of dissipation is antagonistic to that described in the previous section: instead of stabilizing the system against the periodic drive, it actually generates a parametric resonance that would not take place in the corresponding closed system.
Furthermore, the nature of this instability is strikingly different from the conventional parametric instability that appeared in Sec.~\ref{sec:DDPTs_in_TLLs}:
in the present case, only the left-moving quasiparticles (in the rotating time-independent frame) are exponentially amplified.
We therefore dub this phenomenon a \textit{chiral parametric instability}. 
Importantly, this instability emerges from the interplay between the drive and the chiral dissipation: in the absence of dissipation, the system is in a stable phase where quasiparticles are not amplified, and in the absence of Floquet drive there is no coupling between right and left movers, and thus left-moving quasiparticles are unaffected by the dissipation.
More precisely, the chiral instability arises as a consequence of the following equality:
\begin{equation}
\label{eq:chiralcondition1}
\Im(\bar{\Lambda}_{q}^{+}) = -\Im(\bar{\Lambda}_{q}^{-}),
\end{equation}
which we numerically observe to always hold in the chiral unstable phase.
Combined with the observation that \eqref{eq:stabcondchiral} is never satisfied, this equality implies that the Liouvillian spectrum~\eqref{eq:spectrumchiral} has eigenvalues with a positive real part for the left movers, while the eigenvalues always have a nonpositive real part for the right movers, as illustrated in Figs.~\ref{fig:chiral_spectrum}(c) and~\ref{fig:chiral_spectrum}(d).

However, when the driving parameters $(A, \omega/q)$ are chosen to be in the unstable phase in the absence of dissipation, the chiral dissipation is unable to ever fully stabilize the TLL, as opposed to the case in Sec.~\ref{sec:DDPTs_in_TLLs}.
Here, the instability for a given mode $q$ arises as a consequence of the equality
\begin{equation}
\label{eq:chiralcondition2}
\Im(\bar{\Lambda}_{q}^{+}) = \Im(\bar{\Lambda}_{q}^{-}),
\end{equation}
which we numerically observe to hold whenever \eqref{eq:chiralcondition1} is not satisfied.
It follows that the spectrum~\eqref{eq:spectrumchiral} has a positive real part for both right and left movers as long as~\eqref{eq:stabcondchiral} is satisfied, as shown in Figs.~\ref{fig:chiral_spectrum}(a) and~\ref{fig:chiral_spectrum}(b).
This means that both chiralities are equally amplified, which we refer to as a \emph{symmetric parametric instability}.
Indeed, in this case, the real part of the spectrum looks the same as that of \eqref{symmetric_spectrum} but with $\bar{\gamma}$ replaced by $\bar{\gamma}/2$ and $\bar{\Lambda}_{q}$ by $\bar{\Lambda}_{q}^{+} + \bar{\Lambda}_{q}^{-}$.

Our results show that there is a new type of phase transition within the unstable region between a conventional symmetric parametric instability, implied by~\eqref{eq:chiralcondition2}, and a chiral one, implied by~\eqref{eq:chiralcondition1}, when increasing the chiral dissipation $\bar{\gamma}$.
Since our numerical calculations indicate that $|\Im(\bar{\Lambda}_{q}^{+})|$ and $|\Im(\bar{\Lambda}_{q}^{-})|$ are equal (for our drives), either $\Im(\bar{\Lambda}_{q}^{+}) + \Im(\bar{\Lambda}_{q}^{-})$ or $\Im(\bar{\Lambda}_{q}^{+}) - \Im(\bar{\Lambda}_{q}^{-})$ would serve as an ``order parameter'' for this new transition.
The first choice as a function of $\gamma$ is plotted in Figs.~\ref{fig:chiral_phase}(c) and~\ref{fig:chiral_phase}(d), showing sharp jumps from a nonzero value to zero at the phase transition.
Moreover, Fig.~\ref{fig:chiral_phase}(a) shows the same combination plotted for a fixed $\gamma > 0$ but for different driving parameters $(A, \omega/q)$, which can be interpreted as a phase diagram for the chirally dissipated system.

We close this section by highlighting an analogy between the chiral instability and the non-Hermitian skin effect \cite{PhysRevLett.77.570, PhysRevLett.121.086803}.
To see this, we can derive an effective non-Hermitian Hamiltonian from \eqref{eq:timeindepliouvchiral} following a quantum-jump approach \footnote{The effective Hamiltonian is obtained as $\tilde{H}_{\textrm{eff}} = \tilde{H}_{\textrm{stat}} - i \sum_j \gamma_j L_j^\dagger L_j\pdag /2$, where $\tilde{H}_{\textrm{stat}}$ is the Hamiltonian in the unitary part of the time-independent Liouvillian, $L_j$ denote the jump operators in the time-independent dissipator, and $\gamma_j$ are the corresponding couplings.}
\cite{Daley_2014}:
\begin{multline}
\label{H_eff}
\tilde{H}_{\textrm{eff}}
= \sum_{q>0}q \biggl[
  \Re (\bar{\Lambda}_q^+) a_q^{\dagger} a_q\pdag
  + \Re(\bar{\Lambda}_q^-) \bar{a}_q^{\dagger} \bar{a}_q\pdag \\
  + i \Bigl( \Im(\bar{\Lambda}_q^+)-\frac{\bar{\gamma}}{4} \Bigr) a_q^{\dagger} a_q\pdag
  + i \Bigl( \Im(\bar{\Lambda}_q^-)-\frac{\bar{\gamma}}{4} \Bigr) \bar{a}_q^{\dagger} \bar{a}_q\pdag
\biggr].
\end{multline}
In the chiral unstable phase, the condition  \eqref{eq:chiralcondition1} implies an imbalance between the contributions from the last two terms in \eqref{H_eff}.
On the other hand, for the symmetric parametric instability, \eqref{eq:chiralcondition2} is instead satisfied, which implies that the last two terms in \eqref{H_eff} contribute equally.
Since the observed chiral imbalance breaks parity symmetry, this suggests that the system undergoes a transition similar to that of the non-Hermitian skin effect.

%============================================================================
\section{Conclusion}
\label{sec:Conclusion}
%============================================================================

In this work, we showed that the interplay between Floquet drive and dissipation in TLLs may lead to (i) a stabilization of the driven TLL and a suppression of the many-body parametric resonance involving infinitely many bosonic modes, and (ii) dynamical phase transitions between symmetric and chiral parametric instabilities.
We expect that the chiral parametric instability may arise in more generic driven-dissipative scenarios, where dissipation is asymmetric instead of purely chiral \cite{PhysRevLett.131.256602}.
In particular, candidates for potential experimental realizations of this phase include quantum point contacts between two quantum Hall edges, where only one of the two edges is coupled to a bath.
However, the resulting coupling between quantum Hall edges would be spatially local, as opposed to the one studied in the present work.
While one may directly generalize our approach to a time-dependent $\gamma_{t}(q)$ that depends on $q$ so that the dissipation strength $q\gamma_{t}$ is not necessarily linear in $q$, stemming from that different momentum modes are decoupled, we leave the study of such setups for future work.

As pointed out in the previous section, the effective non-Hermitian Hamiltonian in the chiral unstable phase bears similarities with models of the non-Hermitian skin effect, such as the Hatano-Nelson model~\cite{PhysRevLett.77.570}.
It would be desirable to further explore the link between the chiral instability and the Liouvillian skin effect~\cite{PhysRevLett.127.070402}, e.g., by studying spectral flow in the spectrum of the time-independent Liouvillian.
Furthermore, this could draw connections with other driven bosonic systems that have been shown to host non-Hermitian skin effect, such as the bosonic Kitaev chain~\cite{PhysRevX.8.041031, PhysRevB.105.064302}.

Finally, the approach followed in the present work relies on a finite-dimensional algebra of superoperators, which could presumably be generalized beyond Floquet drives to quasiperiodic~\cite{PhysRevLett.120.070602, PhysRevResearch.2.033461, PhysRevLett.126.040601} or even random drives in the presence of dissipation, in which case the stabilization of all the modes is an open question. 

%============================================================================

\begin{acknowledgments}
We thank Apoorv Tiwari for collaboration at an early stage of this project, Stefano Scopa for exchanges concerning Floquet Liouvillians, Tian-Hua Yang for pointing out ideas related to the quantum jump approach, and Ramasubramanian Chitra for inspiring discussions.
B.L.\ acknowledges financial support from the Swiss National Science Foundation (Postdoc.Mobility Grant No.~214461).
P.M.\ acknowledges financial support from the Wenner-Gren Foundations (Grant No.~FT2022-0002).
S.R.\ is supported by a Simons Investigator Grant from the Simons Foundation (Award No.~566116).
\end{acknowledgments}

\textit{Data Availability.}
The data that support the findings of this article are openly available \cite{Drivendissipative}. 

\vspace{-4mm}

%================================================================

% \bibliography{references_ddQLs}

%apsrev4-2.bst 2019-01-14 (MD) hand-edited version of apsrev4-1.bst
%Control: key (0)
%Control: author (8) initials jnrlst
%Control: editor formatted (1) identically to author
%Control: production of article title (0) allowed
%Control: page (0) single
%Control: year (1) truncated
%Control: production of eprint (0) enabled
%

%================================================================
\onecolumngrid
\clearpage
%================================================================

\renewcommand{\theequation}{S\arabic{equation}}
\renewcommand{\thesection}{\Alph{section}}
\setcounter{section}{0}
\setcounter{equation}{0}

%================================================================

\begin{center}
{\Large Supplemental Material}
\end{center}
\csname phantomsection\endcsname%
\addcontentsline{toc}{part}{Supplemental Material}%
%================================================================

This supplemental material contains a number of appendices with technical details for the results in the main text.

%================================================================
\section{Construction of the rotating frame for the symmetric dissipation}
\label{app:rotating_frame}
%================================================================

Our starting point is the time-dependent Liouvillian $\mathcal{L}_t$ in \eqref{eq:masterequation}--\eqref{eq:dissipationsym}.
We first observe that this Liouvillian is built from seven independent superoperators:
$H_{q, 0}$, $H_{q, 1}$, $H_{q, 2}$, $M_{q, 1} = D_{q, 1}+\bar{D}_{q, 1}$, $M_{q, 2} = D_{q, 2}+\bar{D}_{q, 2}$, $M_{q, 3} = D_{q, 3}+\bar{D}_{q, 3}$, and $M_{q, 4} = D_{q, 4}+\bar{D}_{q, 4}$.
These generate a closed algebra, and their commutation relations are summarized in Table~\ref{tab:comm7d}.

In principle, the rotating-frame superoperator $W(t)$ can be expressed as a product of exponentials of the above generators.
However, handling all seven simultaneously can be cumbersome.
To make progress, we make use of the Levi decomposition of the algebra
into the semidirect sum of the semisimple subalgebra generated by $\{ H_{q, 0}, H_{q, 1}, H_{q, 2} \}$ and the solvable ideal generated by $\{ M_{q, 1}, M_{q, 2}, M_{q, 3}, M_{q, 4} \}$.
This enables us to process the action of  $W(t)$ incrementally, applying it layer by layer across each component.

Lastly, we note that the semisimple subalgebra is precisely the $\su(1,1)$ algebra corresponding to the unitary part of the Liouvillian, while the ideal makes up the dissipative part.

\begin{table}[!ht]
\centering
\resizebox{0.6\textwidth}{!}{%
\begin{tabular}{c"c|c|c|c|c|c|c|} 
    & $H_0$ & $H_1$ & $H_2$ & $M_1$ & $M_2$ & $M_3$ & $M_4$ \\ \thickhline\Tstrut
    $H_0$ & 0 & $2iH_1$ & $-2iH_2$ & 0 & 0 & $-2iM_3$ & $2iM_4$ \\ \hline\Tstrut
    $H_1$ & $-2iH_1$ & 0 & $-4iH_0$ & $-2iM_4$ & $-2iM_4$ & $-2i(M_1+M_2)$ & 0 \\ \hline\Tstrut
    $H_2$ & $2iH_2$ & $4iH_0$ & 0 & $2iM_3$ & $2iM_3$ & 0 & $2i(M_1+M_2)$ \\ \hline\Tstrut
    $M_1$ & 0 & $2iM_4$ & $-2iM_3$ & 0 & $-(M_1+M_2)$ & $-M_3$ & $-M_4$ \\ \hline\Tstrut
    $M_2$ & 0 & $2iM_4$ & $-2iM_3$ & $M_1+M_2$ & 0 & $M_3$ & $M_4$ \\ \hline\Tstrut
    $M_3$ & $2iM_3$ & $2i(M_1+M_2)$ & 0 & $M_3$ & $-M_3$ & 0 & 0 \\ \hline\Tstrut
    $M_4$ & $-2iM_4$ & 0 & $-2i(M_1+M_2)$ & $M_4$ & $-M_4$ & 0 & 0 \\ \hline
\end{tabular}%
}%
\caption{Commutators $[A,B]$ for the 7-dimensional algebra for elements $A$ and $B$ in the corresponding row and column.
(The momentum label $q$ is omitted.)}
\label{tab:comm7d}
\end{table}

\begin{table}[!ht]
\centering
\resizebox{\textwidth}{!}{%
\begin{tabular}{c"c|c|c|c|c|c|c|} 
    & $H_0$ & $H_1$ & $H_2$ & $M_1$ & $M_2$ & $M_3$ & $M_4$ \\ \thickhline\Tstrut
    $H_0$ & $H_0$ & $e^{2it}H_1$ & $e^{-2it} H_2$ & $M_1$ & $M_2$ & $e^{-2it}M_3$ & $e^{2it}M_4$ \\ \hline\Tstrut
    $H_1$ & $H_0-2itH_1$ & $H_1$ & $H_2-4itH_0-4t^2H_1$ & $M_1-2itM_4$ & $M_2-2itM_4$ & $M_3-2it(M_1+M_2)-4t^2M_4$ & $M_4$ \\ \hline\Tstrut
    $H_2$ & $H_0+2itH_2$ & $H_1+4itH_0-4t^2H_2$ & $H_2$ & $M_1+2itM_3$ & $M_2+2itM_3$ & $M_3$ & $M_4+2it(M_1+M_2)-4t^2 M_3$ \\ \hline\Tstrut
    $M_1$ & $H_0$ & $H_1-2i(e^{-t}-1)M_4$ & $H_2+2i(e^{-t}-1)M_3$ & $M_1$ & $(e^{-t}-1)M_1+e^{-t}M_2$ & $e^{-t}M_3$ & $e^{-t}M_4$ \\ \hline\Tstrut
    $M_2$ & $H_0$ & $H_1+2i(e^t-1)M_4$ & $H_2-2i(e^t-1)M_3$ & $e^tM_1+(e^t-1)M_2$ & $M_2$ & $e^{t}M_3$ & $e^t M_4$ \\ \hline\Tstrut
    $M_3$ & $H_0+2itM_3$ & $H_1+2it(M_1+M_2)$ & $H_2$ & $M_1+tM_3$ & $M_2-tM_3$ & $M_3$ & $M_4$ \\ \hline\Tstrut
    $M_4$ & $H_0-2itM_4$ & $H_1$ & $H_2-2it(M_1+M_2)$ & $M_1+tM_4$ & $M_2-tM_4$ & $M_3$ & $M_4$ \\ \hline
\end{tabular}%
}%
\caption{Adjoint actions $e^{tA} B e^{-tA}$ for the 7-dimensional algebra for elements $A$ and $B$ in the corresponding row and column.
(The momentum label $q$ is omitted.)}
\label{tab:BCH7d}    
\end{table}

The superoperator $W(t)$ that implements the transformation \eqref{eq:cL_tilde} can now be decomposed as $W(t) = V(t) U(t)$ with $U(t) = \prod_{q>0} U_q(t) = \prod_{q>0}  e^{f_{q, 0} H_{q, 0}} e^{f_{q, 1} H_{q, 1}} e^{f_{q, 2} H_{q, 2}}$ and $V(t) = \prod_{q>0} V_q(t) = e^{g_{q, 1} M_{q, 1}} e^{g_{q, 2} M_{q, 2}} e^{g_{q, 3} M_{q, 3}} e^{g_{q, 4} M_{q, 4}}$.
The effective Liouvillian defined in \eqref{eq:cL_tilde} then becomes
\begin{equation}
\tilde{\mathcal{L}}
= (\partial_t V)V^{-1} + V \Bigl[ (\partial_t U)U^{-1}+ U \mathcal{L}_t U^{-1} \Bigr] V^{-1}.
\end{equation}
We can evaluate the right-hand side using the adjoint actions in Table~\ref{tab:BCH7d}, obtained with the help of the Baker–Campbell–Hausdorff formula.
First, starting with the unitary contribution to $W(t)$, we get
\begin{equation}
(\partial_t U)U^{-1} + U \mathcal{L}_t U^{-1}
= \sum_{q>0} \Bigl[ B_{q, 0}(t) H_{q, 0} + B_{q, 1}(t) H_{q, 1} + B_{q, 2}(t) H_{q, 2} + \tilde{\mathcal{D}}_{q, t} \Bigr]
\end{equation}
with
\begin{equation}
\left\{
\begin{aligned}
    B_{q, 0}(t)
    & = -4i f_{q, 1} e^{2if_{q, 0}} B_{q, 2}(t) + \dot{f}_{q, 0} + 2iq\lambda_t f_{q, 2} + q\epsilon_t, \\
    B_{q, 1}(t)
    & = -4f_{q, 1}^2 e^{4if_{q, 0}} B_{q, 2}(t) + e^{2if_{q, 0}} \Bigl( \dot{f}_{q, 1} + \frac{q\lambda_t}{2} + 4f_{q, 1} f_{q, 2}  q\lambda_t -2iq \epsilon_t f_{q, 1} \Bigr), \\
    B_{q, 2}(t)
    & = e^{-2if_{q, 0}} \Bigl( \dot{f}_{q, 2} + \frac{q\lambda_t}{2} + 2i q\epsilon_t f_{q, 2} - 2 q\lambda_t f_{q, 2}^2 \Bigr), \\
\end{aligned}
\right.
\end{equation}
where 
\begin{equation}
\tilde{\mathcal{D}}_{q, t}
= \gamma_t q \Bigl[ (\mathfrak{n}_{q, t}+1) M_{q, 1} + \mathfrak{n}_{q, t} M_{q, 2} - \mathfrak{m}_{q, t} M_{q, 3} - \mathfrak{m}'_{q, t} M_{q, 4} \Bigr]
\end{equation}
with
\begin{equation}
\left\{
\begin{aligned}
    \mathfrak{n}_{q, t}
    & = - \frac{1}{2} + \Bigl( \mathfrak{n}_{q}+\frac{1}{2} \Bigr) (1+8f_{q, 1}f_{q, 2}) + 2if_{q, 1} \mathfrak{m}_{q} - 2if_{q, 2} (1+4f_{q, 1}f_{q, 2}) \mathfrak{m}_{q}, \\
    \mathfrak{m}_{q, t}
    & = e^{-2if_{q, 0}} \Bigl[ \mathfrak{m}_{q}-4i f_{q, 2} \Bigl( \mathfrak{n}_{q}+\frac{1}{2} \Bigr) - 4 f_{q, 2}^2 \mathfrak{m}_{q} \Bigr], \\
    \mathfrak{m}'_{q, t}
    & = e^{2if_{q, 0}} \Bigl[ \mathfrak{m}_{q} (1+4f_{q, 1} f_{q, 2})^2 + 4i f_{q, 1} (1+ 4f_{q, 1}f_{q, 2}) \Bigl( \mathfrak{n}_{q}+\frac{1}{2} \Bigr) -4 f_{q, 1}^2 \mathfrak{m}_{q} \Bigr].\\
\end{aligned}
\right.
\end{equation}
By choosing $f_{q, 1}$ and $f_{q, 2}$ so that $B_{q, 1}(t) = B_{q, 2}(t) = 0$ and $f_{q, 0} = q \int_0^t \dd t'\, \bigl[ \bar{\Lambda}_q - \Lambda_q(t') \bigr]$ with $\Lambda_q(t) = \epsilon_t + 2i \lambda_t f_{q, 2}$ and $\bar{\Lambda}_q = \frac{1}{T} \int_0^T \dd t\, \Lambda_q(t)$ as in \eqref{eq:Lambdaq}, we obtain
\begin{equation}
(\partial_t U)U^{-1} + U \mathcal{L}_t U^{-1}
= \sum_{q>0} \Bigl( q \bar{\Lambda}_q H_{q, 0} + \tilde{\mathcal{D}}_{q, t} \Bigr).
\end{equation}
Second, including the dissipative contribution, we get
\begin{equation}
\tilde{\mathcal{L}}
= \sum_{q>0} \Bigl( q \bar{\Lambda}_q H_{q, 0} + A_{q, 1}(t) M_{q, 1} + A_{q, 2}(t) M_{q, 2} + A_{q, 3}(t) M_{q, 3} + A_{q, 4}(t) M_{q, 4} \Bigr)
\end{equation}
with
\begin{equation}
\left\{
\begin{aligned}
    A_{q, 1}(t)
    & = A_{q, 2}(t) + \gamma_t q + \dot{g}_{q, 1} -\dot{g}_{q, 2}, \\
    A_{q, 2}(t)
    & = e^{-g_{q, 1}} (\dot{g}_{q, 2}-\gamma_t q) + e^{-g_{q, 1}+g_{q, 2}} \gamma_t q (\mathfrak{n}_{q, t}+1), \\
    A_{q, 3}(t)
    & = e^{-g_{q, 1}+g_{q, 2}} \bigl[ \dot{g}_{q, 3} +(\gamma_t q +2i \bar{\Lambda}_q q)g_{q, 3} - \gamma_t q \mathfrak{m}_{q, t} \bigr], \\
    A_{q, 4}(t)
    & = e^{-g_{q, 1}+g_{q, 2}} \bigl[ \dot{g}_{q, 4}+(\gamma_t q-2i \bar{\Lambda}_q q) g_{q, 4} - \gamma_t q \mathfrak{m}'_{q, t} \bigr]. \\
\end{aligned}
\right.
\end{equation}
In order for the effective Liouvillian to be time-independent, we set
\begin{equation}
\left\{ 
\begin{aligned}
    & g_{q, 1} = g_{q, 2} + q \int_0^t \dd t'\, (\bar{\gamma}-\gamma_{t'}), \\
    & \dot{g}_{q, 2}-\gamma_t q + e^{g_{q, 2}} \gamma_t q (\mathfrak{n}_{q, t}+1) = 0, \\
    & \dot{g}_{q, 3}+(\gamma_t q+ 2i \bar{\Lambda}_q q)g_{q, 3}-\gamma_t q \mathfrak{m}_{q, t} = 0, \\
    & \dot{g}_{q, 4}+(\gamma_t q-2i \bar{\Lambda}_q q) g_{q, 4}- \gamma_t q \mathfrak{m}'_{q, t} = 0 \vphantom{\int_0^t} \\
\end{aligned}
\right.
\end{equation}
with $\bar{\gamma}$ given in \eqref{eq:gammaBar}.
This finally implies
\begin{equation}
\tilde{\mathcal{L}}
= \sum_{q>0} q \Bigl( \bar{\Lambda}_q H_{q, 0} + \bar{\gamma} M_{q, 1} \Bigr),
\end{equation}
and using that $M_{q, 1} = D_{q, 1}+\bar{D}_{q, 1}$, we obtain the desired form \eqref{eq:effective_Liouvillian} of the effective Liouvillian.

%================================================================
\section{Spectrum of the time-independent Liouvillian}
\label{app:HOspectrum}
%================================================================

In this appendix, we review the use of vectorization to derive the spectrum for the time-independent Liouvillian in \eqref{eq:effective_Liouvillian}.
The Liouvillian in \eqref{eq:timeindepliouvchiral} can be treated similarly.
Notably, as both these Liouvillians are block-diagonal with respect to both momentum $q$ and chirality, we can simplify our analysis by considering only a single sector, say, of right movers with momentum $q > 0$.

Thus, in what follows, it suffices to study a time-independent Liouvillian of the form
\begin{equation}
\tilde{\mathcal{L}}
= \bar{\Lambda} H_0 + \bar{\gamma} D_1
\end{equation}
with the superoperators
\begin{equation}
H_0
= -i [a^\dagger a, \bullet],
\qquad
D_1
= a \bullet a^\dagger -\frac{1}{2} \{a^\dagger a, \bullet\},
\end{equation}
where we dropped all momentum subscripts for simplicity.
The Liouvillian acts on density matrices $\rho$, which we can formally express as $\rho = \sum_{n, m} \rho_{n,m} |n\rangle\langle m|$ in some basis $\{ |n\rangle \}$ labeled by the number $n$ of bosons in each state.
As is commonplace, we may vectorize $\rho$ as $\sum_{n, m} \rho_{n,m} |m\rangle \otimes |n\rangle$.
In terms of superoperators $A \bullet$ and $\bullet B$, this corresponds to the vectorization rules
\begin{equation}
A \bullet \to I \otimes A,
\qquad
\bullet B \to B^T \otimes I,
\end{equation}
where ${}^T$ denotes transpose and we used the same symbol for the usual operators on the right-hand sides to simplify notation, with the distinction that $A$ and $B^T$ act on separate Hilbert spaces.
Using these rules, we can identify the superoperators in the Liouvillian as follows:
\begin{equation}
\begin{aligned}
H_0 & = -i I \otimes a^\dagger a + i (a^\dagger a)^T \otimes I, \\
D_1 & = (a^\dagger)^T \otimes a -\frac{1}{2} I\otimes a^\dagger a -\frac{1}{2} (a^\dagger a)^T \otimes I.
\end{aligned}
\end{equation}
Since the pairs $a$, $a^\dagger$ and $a^T$, $(a^\dagger)^T$ act in separate Hilbert spaces, they are different operators, and we may thus denote the latter as a new pair of bosonic creation and annihilation operators $b^\dagger$, $b$ [which preserves the commutation relations since $1 = a a^\dagger - a^\dagger a$ implies $1 = (a a^\dagger)^T - (a^\dagger a)^T = b b^\dagger - b^\dagger b$].
Omitting to write out the tensor products, we therefore obtain
\begin{equation}
\label{eq:mcL_vectorized}
\tilde{\mathcal{L}}
= \Bigl( -i\bar{\Lambda} - \frac{1}{2}\bar{\gamma} \Bigr) a^\dagger a + \Bigl( i\bar{\Lambda} - \frac{1}{2} \bar{\gamma} \Bigr) b^\dagger b + \bar{\gamma} ba, 
\end{equation}
where $\bigl[ a, a^\dagger \bigr] = 1 = \bigl[b, b^\dagger \bigr]$ and $\bigl[ a, b \bigr] = 0$.

We want to solve the eigenvalue problem $\tilde{\mathcal{L}} \rho = E \rho$ for $\tilde{\mathcal{L}}$ in \eqref{eq:mcL_vectorized} acting on matrices $\rho = \sum_{n,m} \rho_{n,m} |n\rangle_a |m\rangle_b$, where the sums are over nonnegative integers $n$, $m$.
Since
\begin{equation}
\tilde{\mathcal{L}} \sum_{n,m} \rho_{n,m} |n\rangle_a |m\rangle_b
= \sum_{n,m} \biggl( \biggl[ \Bigl( -i\bar{\Lambda} - \frac{1}{2}\bar{\gamma} \Bigr) n + \Bigl( i\bar{\Lambda} - \frac{1}{2}\bar{\gamma} \Bigr) m \biggr] \rho_{n,m} |n\rangle_a |m\rangle_b + \bar{\gamma} \sqrt{nm} \rho_{n,m} |n-1\rangle_a |m-1\rangle_b \biggr)
\end{equation}
and using orthogonality, we obtain the following recursion relation:
\begin{equation}
\label{eq:rec_rel}
\bar{\gamma} \sqrt{(n+1)(m+1)} \rho_{n+1,m+1}
= \biggl[ E - \Bigl( -i\bar{\Lambda} -\frac{1}{2}\bar{\gamma} \Bigr) n - \Bigl( i\bar{\Lambda} - \frac{1}{2} \bar{\gamma} \Bigr) m \biggr] \rho_{n,m}.
\end{equation}
Due to the relative powers in $n$, $m$ between the coefficients on the left- and right-hand sides of \eqref{eq:rec_rel}, the recursion relation must terminate in order for $\sum_{n,m} \rho_{n,m} |n\rangle_a |m\rangle_b$ to be convergent.
I.e., $\rho_{N+1,M+1} = 0$ for some pair $(N,M)$, which happens if the corresponding eigenvalue is
\begin{equation}
E_{N,M}
= i\bar{\Lambda} (M-N) - \frac{\bar{\gamma}}{2} (M+N).
\end{equation}
Taken together, these eigenvalues make up one sector of the spectrum in \eqref{symmetric_spectrum}, with $(N, M) = (n_1, m_1)$ for right movers and $(n_2, m_2)$ for left movers, respectively.
The spectrum in \eqref{eq:spectrumchiral} is obtained analogously.

%================================================================
\section{Derivation of the Floquet Liouvillian for the symmetric dissipation}
\label{sec:floquetliouvil}
%================================================================

As is manifest from \eqref{eq:cL_F}--\eqref{eq:rotatingevolution}, the stroboscopic time evolution of our system is captured by the Floquet Liouvillian
\begin{equation}
\mathcal{L}_F(t)
= W(t)^{-1} \tilde{\mathcal{L}}_t W(t)
= U(t)^{-1} V(t)^{-1} \tilde{\mathcal{L}}_t V(t) U(t),
\end{equation}
where $U(t) = \prod_{q>0} U_q(t)$ and $V(t) = \prod_{q>0} V_q(t)$ are given by \eqref{UqVq}.
To obtain an explicit formula, we first start from the time-independent effective Liouvillian \eqref{eq:effective_Liouvillian} and apply $V^{-1} \bullet V$, yielding
\begin{equation}
V(t)^{-1} \tilde{\mathcal{L}}_t V(t)
= \sum_{q>0}  q \Bigl[ \bar{\Lambda}_q H_{q, 0} + \bar{\gamma} e^{-g_{q, 2}} M_{q, 1} + \bar{\gamma}(e^{-g_{q, 2}}-1)M_{q, 2} - g_{q, 3} (\bar{\gamma}+2i\bar{\Lambda}) M_{q, 3} - g_{q, 4} (\bar{\gamma} - 2i\bar{\Lambda}) M_{q, 4} \Bigr].
\end{equation}
By applying $U^{-1} \bullet U$, we finally obtain
\begin{equation}
\label{symmetric_floquet}
\mathcal{L}_F(t)
= \sum_{q>0} q \biggl[ \omega^F_{q, t} H_{q, 0} + \frac{\lambda^F_{q, t}}{2} H_{q, 1} + \frac{\lambda'^F_{q, t}}{2} H_{q, 2} + \bar{\gamma} \Bigl( \bigl[ \mathfrak{n}^F_{q, t}+1 \bigr] M_{q, 1} + \mathfrak{n}^F_{q, t} M_{q, 2} - \mathfrak{m}^F_{q, t} M_{q, 3} - \mathfrak{m}'^F_{q, t} M_{q, 4} \Bigr) \biggr]
\end{equation}
with
\begin{equation}
\left\{
\begin{aligned}
    \omega^F_{q, t}
    & = \bar{\Lambda}_q (1+8f_{q, 1}f_{q, 2}), \\
    \lambda^F_{q, t}
    & = 4if_{q, 1} \bar{\Lambda}_q, \\
    \lambda'^F_{q, t}
    & = -4if_{q, 2} (1+4f_{q, 1}f_{q, 2}) \bar{\Lambda}_q, \\
    \mathfrak{n}^F_{q, t}
    & = - \frac{1}{2} + (1+8f_{q, 1}f_{q, 2}) \Bigl( e^{-g_{q, 2}}-\frac{1}{2} \Bigr) - \frac{2i}{\bar{\gamma}} e^{2if_{q, 0}} g_{q, 3} f_{q, 1}  (1+4f_{q, 1} f_{q, 2}) (\bar{\gamma} + 2i\bar{\Lambda}_q) + \frac{2i}{\bar{\gamma}} e^{-2if_{q, 0}} g_{q, 4} f_{q, 2} (\bar{\gamma} - 2i\bar{\Lambda}_q), \\
    \mathfrak{m}^F_{q, t}
    & = 4i f_{q, 2} (1+4f_{q, 1} f_{q, 2}) \Bigl( e^{-g_{q, 2}}-\frac{1}{2} \Bigr) + \frac{1}{\bar{\gamma}} \Bigl[ -4 e^{-2if_{q, 0}} g_{q, 4} f_{q, 2}^2 (\bar{\gamma}-2i\bar{\Lambda}_q) + e^{2if_{q, 0}} g_{q, 3} (1+4f_{q, 1} f_{q, 2})^2 (\bar{\gamma} + 2i\bar{\Lambda}_q) \Bigr], \\
    \mathfrak{m}'^F_{q, t}
    & = -4if_{q, 1} \Bigl( e^{-g_{q, 2}}-\frac{1}{2} \Bigr) + \frac{1}{\bar{\gamma}} \Bigl[ e^{-2if_{q, 0}} g_{q, 4} (\bar{\gamma} - 2i\bar{\Lambda}_q) - 4e^{2if_{q, 0}} g_{q, 3} f_{q, 1}^2 (\bar{\gamma} + 2i\bar{\Lambda}_q) \Bigr],
\end{aligned}
\right.
\end{equation}
which yields \eqref{cL_F_t0} at time $t_0$.

%================================================================
\section{Equation of motion for correlators}
\label{sec:EOMoperators}
%================================================================

In this appendix, we give the details for the time evolution of bosonic correlators in Sec.~\ref{subsec:te_obs}.
First, we consider the evolution of the ``dressed'' density matrix $\tilde{\rho}_t = e^{(t-t_0)\mathcal{L}_F(t_0)} \rho_{t_0}$.
The equation of motion for this density matrix is
\begin{equation}
\partial_t \tilde{\rho}_t = \mathcal{L}_F (t_0) \rho_{t_0}.
\end{equation}
It then follows that the equations of motion for the correlators take the following form:
\begin{equation}
\left\{
\begin{aligned}
    \partial_t \langle a_q^\dagger a_q\pdag \rangle_{t}
    & = i q\lambda^F_{q, t_0} \langle a_q\pdag \bar{a}_q\pdag \rangle_{t} - iq \lambda'^F_{q, t_0} \langle a_q^\dagger \bar{a}_q^\dagger \rangle_{t} + q\bar{\gamma} \bigl( \mathfrak{n}^F_{q, t_0} - \langle a_q^\dagger a_q\pdag \rangle_{t} \bigr), \\
   \partial_t \langle \bar{a}_q^\dagger \bar{a}_q\pdag \rangle_{t}
    & = i q\lambda^F_{q, t_0} \langle a_q\pdag \bar{a}_q\pdag \rangle_{t} - iq \lambda'^F_{q, t_0} \langle a_q^\dagger \bar{a}_q^\dagger \rangle_{t} + q\bar{\gamma} \bigl( \mathfrak{n}^F_{q, t_0} - \langle \bar{a}_q^\dagger \bar{a}_q\pdag \rangle_{t} \bigr), \\
    \partial_t \langle a_q\pdag \bar{a}_q\pdag \rangle_{t}
    & = -2i q\omega^F_{q, t_0} \langle a_q\pdag \bar{a}_q\pdag \rangle_{t} - i q \lambda'^F_{q, t_0} \bigl( 1 + \langle a_q^\dagger a_q\pdag \rangle_{t} + \langle \bar{a}_q^\dagger \bar{a}_q\pdag \rangle_{t} \bigr) - q \bar{\gamma} \bigl( \langle a_q\pdag \bar{a}_q\pdag \rangle_{t} - \mathfrak{m}^F_{q, t_0} \bigr), \\
    \partial_t \langle a_q^\dagger \bar{a}_q^\dagger \rangle_{t}
    & = 2iq\omega^F_{q, t_0} \langle a_q^\dagger \bar{a}_q^\dagger \rangle_{t} + i q\lambda^F_{q, t_0} \bigl( 1 + \langle a_q^\dagger a_q\pdag \rangle_{t} + \langle \bar{a}_q^\dagger \bar{a}_q\pdag \rangle_{t} \bigr) - q\bar{\gamma} \bigl( \langle a_q^\dagger \bar{a}_q^\dagger \rangle_{t} - \mathfrak{m}'^F_{q, t_0} \bigr),
\end{aligned}
\right.
\end{equation}
where we recall that $\langle \mathcal{O} \rangle_{t} = \Tr \bigl[ \tilde{\rho}_t \mathcal{O} \bigr]$ denotes the expectation value with respect to $\tilde{\rho}_t$.

%================================================================
\section{Construction of the rotating frame for the chiral dissipation}
\label{sec:chira_dissp}
%================================================================

Here, we turn to the chiral coupling between the system and the bath modeled by \eqref{eq:masterequation} together with \eqref{chiral_Liouvillian}.
As stated in the main text, the Liouvillian lies in an 11-dimensional algebra $\mathfrak{g}$ generated by $H_{q, 0}$, $H_{q, 1}$, $H_{q, 2}$, $D_{q, 1}$, $D_{q, 2}$, $\bar{D}_{q, 1}$, $\bar{D}_{q, 2}$, $D_{q, 3}$, $D_{q, 4}$, $\bar{D}_{q, 3}$, and $\bar{D}_{q, 4}$.
In order to apply the strategy of rotating frames outlined in Appendix~\ref{app:rotating_frame}, we first need to find the Levi decomposition of this 11-dimensional algebra.
The result is the semidirect sum
\begin{equation}
\mathfrak{g} = (\mathfrak{g}_+ \oplus \mathfrak{g}_-) \oplus R_{\mathfrak{g}}
\end{equation}
of the semisimple subalgebra $\mathfrak{g}_+ \oplus \mathfrak{g}_-$ comprised of two $\su(1, 1)$ algebras $\mathfrak{g}_{\pm}$ spanned by
$e_{q, 4/1} = \frac{1}{2} H_{q, 0} \pm i(D_{q, 1} + \bar{D}_{q, 2})$,
$e_{q, 5/2} = \frac{1}{2} H_{q, 1} \pm 2iD_{q, 4}$,
$e_{q, 6/3} = \frac{1}{2} H_{q, 2} \pm  2iD_{q, 3}$,
respectively, and the ideal $R_\mathfrak{g}$ spanned by
$v_{q, 1} = D_{q, 1} - \bar{D}_{q, 2}$,
$v_{q, 2} = D_{q, 1} + D_{q, 2}$,
$v_{q, 3} = \bar{D}_{q, 1} + \bar{D}_{q, 2}$,
$v_{q, 4} = D_{q, 3} + \bar{D}_{q, 3}$,
$v_{q, 5} = D_{q, 4}+ \bar{D}_{q, 4}$.
In contrast to the case of a symmetric dissipation, where the semisimple algebra corresponded only to the unitary part of the Liouvillian and the ideal to the dissipative part, the algebraic decomposition for the chiral dissipation exhibits a more complicated structure.
Indeed, in the present case, the semisimple algebra has both unitary and dissipative contributions, whereas the ideal remains purely dissipative.
The commutation relations for the 11-dimensional algebra are given in Table~\ref{tab:comm11d}, and the necessary adjoint actions are given in Table~\ref{tab:BCH11d}, again obtained with the help of the Baker–Campbell–Hausdorff formula.

\begin{table}[!ht]
\centering
\resizebox{0.6\textwidth}{!}{%
\begin{tabular}{c"c|c|c|c|c|c|c|c|c|c|c|}  
    & $e_1$ & $e_2$ & $e_3$ & $e_4$ & $e_5$ & $e_6$ & $v_1$ & $v_2$ & $v_3$ & $v_4$ & $v_5$ \\ \thickhline\Tstrut
    $e_1$ & $0$ & $2ie_2$ & $-2ie_3$ & $0$ & $0$ & $0$ & $0$ & $iv_2$ & $-iv_3$ & $-iv_4$ & $iv_5$ \\ \hline\Tstrut
    $e_2$ & $-2ie_2$ & $0$ & $-4ie_1$ & $0$ & $0$ & $0$ & $0$ & $0$ & $-2iv_5$ & $-2iv_2$ & $0$ \\ \hline\Tstrut
    $e_3$ & $2ie_3$ & $4ie_1$ & $0$ & $0$ & $0$ & $0$ & $0$ & $2iv_4$ & $0$ & $0$ & $2iv_3$ \\ \hline\Tstrut
    $e_4$ & $0$ & $0$ & $0$ & $0$ & $2ie_5$ & $-2ie_6$ & $0$ & $-iv_2$ & $iv_3$ & $-iv_4$ & $iv_5$ \\ \hline\Tstrut
    $e_5$ & $0$ & $0$ & $0$ & $-2ie_5$ & $0$ & $-4ie_4$ & $0$ & $-2iv_5$ & $0$ & $-2iv_3$ & $0$ \\ \hline\Tstrut
    $e_6$ & $0$ & $0$ & $0$ & $2ie_6$ & $4ie_4$ & $0$ & $0$ & $0$ & $2iv_4$ & $0$ & $2iv_2$ \\ \hline\Tstrut
    $v_1$ & $0$ & $0$ & $0$ & $0$ & $0$ & $0$ & $0$ & $-v_2$ & $-v_3$ & $-v_4$ & $-v_5$ \\ \hline\Tstrut
    $v_2$ & $-iv_2$ & $0$ & $-2iv_4$ & $iv_2$ & $2iv_5$ & $0$ & $v_2$ & $0$ & $0$ & $0$ & $0$ \\ \hline\Tstrut
    $v_3$ & $iv_3$ & $2iv_5$ & $0$ & $-iv_3$ & $0$ & $-2iv_4$ & $v_3$ & $0$ & $0$ & $0$ & $0$ \\ \hline\Tstrut
    $v_4$ & $iv_4$ & $2iv_2$ & $0$ & $iv_4$ & $2iv_3$ & $0$ & $v_4$ & $0$ & $0$ & $0$ & $0$ \\ \hline\Tstrut
    $v_5$ & $-iv_5$ & $0$ & $-2iv_3$ & $-iv_5$ & $0$ & $-2iv_2$ & $v_5$ & $0$ & $0$ & $0$ & $0$ \\ \hline
\end{tabular}%
}%
\caption{Commutators $[A,B]$ for the 11-dimensional algebra for elements $A$ and $B$ in the corresponding row and column.
It is manifest that the subalgebras generated by $\{ e_1, e_2, e_3 \}$ and $\{ e_4, e_5, e_6 \}$, respectively, are two independent copies of $\su(1,1)$.
(The momentum label $q$ is omitted.)}
\label{tab:comm11d}    
\end{table}

\begin{table}[!ht]
\centering
\resizebox{\textwidth}{!}{%
\begin{tabular}{c"c|c|c|c|c|c|c|c|c|c|c|}  
    & $e_1$ & $e_2$ & $e_3$ & $e_4$ & $e_5$ & $e_6$ & $v_1$ & $v_2$ & $v_3$ & $v_4$ & $v_5$ \\ \thickhline\Tstrut
    $e_1$ & $e_1$ & $e^{2it}e_2$ & $e^{-2it}e_3$ & $e_4$ & $e_5$ & $e_6$ & $v_1$ & $e^{it}v_2$ & $e^{-it}v_3$ & $e^{-it}v_4$ & $e^{it}v_5$ \\ \hline\Tstrut
    $e_2$ & $e_1-2ite_2$ & $e_2$ & $e_3-4ite_1-4t^2e_2$ & $e_4$ & $e_5$ & $e_6$ & $v_1$ & $v_2$ & $v_3-2itv_5$ & $v_4-2itv_2$ & $v_5$ \\ \hline\Tstrut
    $e_3$ & $e_1+2ite_3$ & $e_2+4ite_1-4t^2e_3$ & $e_3$ & $e_4$ & $e_5$ & $e_6$ & $v_1$ & $v_2+2itv_4$ & $v_3$ & $v_4$ & $v_5+2itv_3$ \\ \hline\Tstrut
    $e_4$ & $e_1$ & $e_2$ & $e_3$ & $e_4$ & $e^{2it}e_5$ & $e^{-2it}e_6$ & $v_1$ & $e^{-it}v_2$ & $e^{it}v_3$ & $e^{-it}v_4$ & $e^{it}v_5$ \\ \hline\Tstrut
    $e_5$ & $e_1$ & $e_2$ & $e_3$ & $e_4-2ite_5$ & $e_5$ & $e_6-4ite_4-4t^2e_5$ & $v_1$ & $v_2-2itv_5$ & $v_3$ & $v_4-2itv_3$ & $v_5$ \\ \hline\Tstrut
    $e_6$ & $e_1$ & $e_2$ & $e_3$ & $e_4+2ite_6$ & $e_5+4ite_4-4t^2e_6$ & $e_6$ & $v_1$ & $v_2$ & $v_3+2itv_4$ & $v_4$ & $v_5+2itv_2$ \\ \hline\Tstrut
    $v_1$ & $e_1$ & $e_2$ & $e_3$ & $e_4$ & $e_5$ & $e_6$ & $v_1$ & $e^{-t}v_2$ & $e^{-t}v_3$ & $e^{-t}v_4$ & $e^{-t}v_5$ \\ \hline\Tstrut
    $v_2$ & $e_1-itv_2$ & $e_2$ & $e_3-2itv_4$ & $e_4+itv_2$ & $e_5+2itv_5$ & $e_6$ & $v_1+tv_2$ & $v_2$ & $v_3$ & $v_4$ & $v_5$ \\ \hline\Tstrut
    $v_3$ & $e_1+itv_3$ & $e_2+2itv_5$ & $e_3$ & $e_4-itv_3$ & $e_5$ & $e_6-2itv_4$ & $v_1+tv_3$ & $v_2$ & $v_3$ & $v_4$ & $v_5$ \\ \hline\Tstrut
    $v_4$ & $e_1+itv_4$ & $e_2+2itv_2$ & $e_3$ & $e_4+itv_4$ & $e_5+2itv_3$ & $e_6$ & $v_1+tv_4$ & $v_2$ & $v_3$ & $v_4$ & $v_5$ \\ \hline\Tstrut
    $v_5$ & $e_1-itv_5$ & $e_2$ & $e_3-2itv_3$ & $e_4-itv_5$ & $e_5$ & $e_6-2itv_2$ & $v_1+tv_5$ & $v_2$ & $v_3$ & $v_4$ & $v_5$ \\ \hline
\end{tabular}%
}%
\caption{Adjoint actions $e^{tA} B e^{-tA}$ for the 11-dimensional algebra for elements $A$ and $B$ in the corresponding row and column.
(The momentum label $q$ is omitted.)}
\label{tab:BCH11d}
\end{table}

In terms of the new basis, the Liouvillian can be expressed as
\begin{equation}
\mathcal{L}_t
= \sum_{q>0} q \biggl[ \Bigl( \epsilon_t-\frac{\gamma_t}{4i} \Bigr) e_{q, 1} + \Bigl( \epsilon_t+\frac{\gamma_t}{4i} \Bigr) e_{q, 4} + \frac{\lambda_t}{2}(e_{q, 2} + e_{q, 5}) + \frac{\lambda_t}{2} (e_{q, 3}+e_{q, 6}) + \gamma_t \Bigl( \frac{1}{2} v_{q, 1} + \mathfrak{n}_{q} v_{q, 2} \Bigr) \biggr].
\end{equation}
Following a similar procedure as for the case with a symmetric dissipation, we decompose the superoperator $W(t) = U(t) V(t)$ into semisimple $U(t)$ and ideal $V(t)$ contributions.
First, we rotate using $U(t) = U_1(t) U_2(t)$ with
$U_1(t) =\prod_{q>0} U_{q,1}(t)=\prod_{q>0} e^{f_{q, 0}^{-} e_{q, 1}} e^{f_{q, 1}^{-} e_{q, 2}} e^{f_{q, 2}^{-} e_{q, 3}}$ and $U_2(t) = \prod_{q>0} U_{q, 2}(t) = \prod_{q>0} e^{f_{q, 0}^{+} e_{q, 4}} e^{f_{q, 1}^{+} e_{q, 5}} e^{f_{q,2}^{+} e_{q, 6}}$.
This yields
\begin{multline}
(\partial_t U_1)U_1^{-1} + (\partial_t U_2) U_2^{-1} + U_1 U_2 \mathcal{L}_t U_2^{-1} U_1^{-1} \\
= \sum_{q>0} \Bigl( B_{q, 1}(t) e_{q, 1} + B_{q, 2}(t) e_{q, 2} + B_{q, 3}(t) e_{q, 3} + B_{q, 4}(t) e_{q, 4} + B_{q, 5}(t) e_{q, 5} + B_{q, 6}(t) e_{q, 6} + \tilde{\mathcal{D}}_{q, t} \Bigr)
\end{multline}
with 
\begin{equation}
\left\{
\begin{aligned}
    B_{q, 1}(t)
    & = -4if_{q, 1}^{-} e^{2if_{q, 0}^{-}} B_{q, 3}(t) + \Bigl[ \dot{f}_{q, 0}^{-} + 2iq\lambda_t f_{q, 2}^{-} + q \Bigl( \epsilon_t - \frac{\gamma_t}{4i} \Bigr) \Bigr], \\
    B_{q, 2}(t)
    & = -4 (f_{q, 1}^{-})^2 e^{4if_{q, 0}^{-}} B_{q, 3}(t) + e^{2if_{q, 0}^{-}} \Bigl[ \dot{f}_{q, 1}^{-} + q\frac{\lambda_t}{2} +4q\lambda_t f_{q, 1}^{-}f_{q, 2}^{-} -2iq \Bigl( \epsilon_t-\frac{\gamma_t}{4i} \Bigr) f_{q, 1}^{-} \Bigr], \\
    B_{q, 3}(t)
    & = e^{-2if_{q, 0}^{-}} \Bigl[ \dot{f}_{q, 2}^{-} + q\frac{\lambda_t}{2} + 2iq \Bigl( \epsilon_t-\frac{\gamma_t}{4i} \Bigr) f_{q, 2}^{-} -2q\lambda_t (f_{q, 2}^{-})^2 \Bigr], \\
    B_{q, 4}(t)
    & = -4if_{q, 1}^{+} e^{2if_{q, 0}^{+}} B_{q, 6}(t) + \Bigl[ \dot{f}_{q, 0}^{+}  + 2iq\lambda_t f_{q, 2}^{+}+ q \Bigl( \epsilon_t+\frac{\gamma_t}{4i} \Bigr) \Bigr], \\
    B_{q, 5}(t)
    & = -4 (f_{q, 1}^{+})^2 e^{4if_{q, 0}^{+}} B_{q, 6}(t) + e^{2if_{q, 0}^{+}} \Bigl[ \dot{f}_{q, 1}^{+} + q\frac{\lambda_t}{2} + 4 q \lambda_t f_{q, 1}^{+} f_{q, 2}^{+} -2i q \Bigl( \epsilon_t+\frac{\gamma_t}{4i} \Bigr) f_{q, 1}^{+} \Bigr], \\
    B_{q, 6}(t)
    & = e^{-2if_{q, 0}^{+}} \Bigl[ \dot{f}_{q, 2}^{+} + q\frac{\lambda_t}{2} + 2i q \Bigl( \epsilon_t+\frac{\gamma_t}{4i} \Bigr) f_{q, 2}^{+}  -2 q \lambda_t (f_{q, 2}^{+})^2 \Bigr].  \\    
\end{aligned}
\right.
\end{equation}
In this case, we choose $f_{q, 0}^{\pm}$, $f_{q, 1}^{\pm}$, and $f_{q, 2}^{\pm}$ so that
\begin{equation}
\left\{
\begin{aligned}
    & \dot{f}_{q, 2}^{-} + q\frac{\lambda_t}{2} + 2iq \Bigl( \epsilon_t-\frac{\gamma_t}{4i} \Bigr) f_{q, 2}^{-} -2q\lambda_t (f_{q, 2}^{-})^2 = 0, \\
    & \dot{f}_{q, 1}^{-} + q\frac{\lambda_t}{2} + 4q\lambda_t f_{q, 1}^{-}f_{q, 2}^{-} -2iq \Bigl( \epsilon_t-\frac{\gamma_t}{4i} \Bigr) f_{q, 1}^{-} =0, \\
    & \dot{f}_{q, 2}^{+} + q\frac{\lambda_t}{2} + 2i q \Bigl( \epsilon_t+\frac{\gamma_t}{4i} \Bigr) f_{q, 2}^{+}  -2 q \lambda_t (f_{q, 2}^{+})^2  = 0, \\
    & \dot{f}_{q, 1}^{+} + q\frac{\lambda_t}{2} + 4 q \lambda_t f_{q, 1}^{+} f_{q, 2}^{+} -2i q \Bigl( \epsilon_t+\frac{\gamma_t}{4i} \Bigr) f_{q, 1}^{+}  =0,\\
    & f_{q, 0}^{-} = q \int_0^t \dd t'\, \bigl( \bar{\Lambda}_{q}^{-} - \Lambda_{q}^{-}(t') \bigr), \qquad \Lambda_{q}^{-}(t) = \Bigl( \epsilon_t - \frac{\gamma_t}{4i} \Bigr) + 2i \lambda_t f_{q, 2}^{-}, \\
    & f_{q, 0}^{+} = q \int_0^t \dd t'\, \bigl( \bar{\Lambda}_{q}^{+} - \Lambda_{q}^{+}(t') \bigr), \qquad \Lambda_{q}^{+}(t) = \Bigl( \epsilon_t +\frac{\gamma_t}{4i} \Bigr) + 2i \lambda_t f_{q, 2}^{+}. \\
\end{aligned}
\right.
\end{equation}
With this choice, the action of the semisimple rotation becomes
\begin{equation}
\begin{aligned}
(\partial_t U_1)U_1^{-1} + (\partial_t U_2) U_2^{-1} + U_1 U_2 \mathcal{L}_t U_2^{-1} U_1^{-1}
& = \sum_{q>0}
    \Bigl( q\bar{\Lambda}_{q}^{-} e_{q, 1} + q\bar{\Lambda}_{q}^{+} e_{q, 4} + \tilde{\mathcal{D}}_{q, t} \Bigr) \\
& = \sum_{q>0}
    \biggl[ \frac{q}{2} \bigl(\bar{\Lambda}_{q}^{+} + \bar{\Lambda}_{q}^{-} \bigr) H_{q, 0} + i q \bigl( \bar{\Lambda}_{q}^{+} -\bar{\Lambda}_{q}^{-} \bigr) (D_{q, 1} + \bar{D}_{q, 2}) + \tilde{\mathcal{D}}_{q, t} \biggr],
\end{aligned}
\end{equation}
where
\begin{equation}
\tilde{\mathcal{D}}_{q, t}
= \gamma_t q \biggl( \frac{1}{2}v_{q, 1} + \mathfrak{n}_{q,t} v_{q, 2} + \mathfrak{n}'_{q, t} v_{q, 3} + \mathfrak{m}_{q, t} v_{q, 4} + \mathfrak{m}'_{q, t} v_{q, 5} \biggr)
\end{equation}
with
\begin{equation}
\left\{
\begin{aligned}
    \mathfrak{n}_{q, t}
    & = \mathfrak{n}_{q} (1+ 4 f_{q, 1}^{-} f_{q, 2}^{-}) e^{-i (f_{q, 0}^{+} - f_{q, 0}^{-})}, \\
    \mathfrak{n}'_{q, t}
    & = 4 f_{q, 1}^{+} f_{q, 2}^{-} \mathfrak{n}_{q} e^{i(f_{q, 0}^{+}-f_{q, 0}^{-})}, \\
    \mathfrak{m}_{q, t}
    & = 2i \mathfrak{n}_{q} f_{q, 2}^{-} e^{-i(f_{q, 0}^{+}+f_{q, 0}^{-})}, \\
    \mathfrak{m}'_{q, t}
    & = -2i \mathfrak{n}_{q} f_{q, 1}^{+} e^{i(f_{q, 0}^{+}+f_{q, 0}^{-})} (1+4f_{q, 1}^{-} f_{q, 2}^{-}). \\
\end{aligned}
\right.
\end{equation}
Second, including the ideal contribution by rotating using
$V(t) = \prod_{q>0} V_q(t) = \prod_{q>0} \prod_{i=1}^5 e^{g_{q, i} v_{q, i}}$,
we obtain
\begin{equation}
\tilde{\mathcal{L}}
= \sum_{q>0}
  \biggl[ \frac{q}{2} \bigl( \bar{\Lambda}_{q}^{+} + \bar{\Lambda}_{q}^{-} \bigr) H_{q, 0} + iq \bigl( \bar{\Lambda}_{q}^{+} - \bar{\Lambda}_{q}^{-} \bigr) (D_{q, 1} + \bar{D}_{q, 2}) + A_{q, 1} (t) v_{q, 1} + A_{q, 2} (t) v_{q, 2} + A_{q, 3} (t) v_{q, 3} + A_{q, 4} (t) v_{q, 4} + A_{q, 5} (t) v_{q, 5} \biggr],
\end{equation}
where
\begin{equation}
\left\{
\begin{aligned}
    A_{q, 1}(t)
    & = \frac{\gamma_t}{2} q + \dot{g}_{q, 1}, \\
    A_{q, 2} (t)
    & = e^{-g_{q, 1}} \Bigl[ \dot{g}_{q, 2} + \gamma_t q \mathfrak{n}_{q, t}+ g_{q, 2} q \Bigl( \frac{1}{2}\gamma_t - i \bar{\Lambda}_{q}^{-} + i\bar{\Lambda}_{q}^{+} \Bigr) \Bigr], \\
    A_{q, 3}(t)
    & = e^{-g_{q, 1}} \Bigl[ \dot{g}_{q, 3} + \gamma_t q\mathfrak{n}'_{q, t} + g_{q, 3} q \Bigl( \frac{1}{2}\gamma_t + i\bar{\Lambda}_{q}^{-} - i \bar{\Lambda}_{q}^{+} \Bigr) \Bigr], \\
    A_{q, 4}(t)
    & = e^{-g_{q, 1}} \Bigl[ \dot{g}_{q, 4} + \gamma_t q \mathfrak{m}_{q, t} + g_{q, 4} q \Bigl( \frac{1}{2}\gamma_t + i\bar{\Lambda}_{q}^{-} + i\bar{\Lambda}_{q}^{+} \Bigr) \Bigr], \\
    A_{q, 5}(t)
    & = e^{-g_{q, 1}} \Bigl[ \dot{g}_{q, 5} + \gamma_t q \mathfrak{m}'_{q, t} + g_{q, 5} q \Bigl( \frac{1}{2}\gamma_t - i\bar{\Lambda}_{q}^{-} - i\bar{\Lambda}_{q}^{+} \Bigr) \Bigr]. \\
\end{aligned}
\right.
\end{equation}
To obtain a time-independent Liouvillian, we set
\begin{equation}
\left\{
\begin{aligned}
    & g_{q, 1} = \frac{q}{2} \int_0^t \dd t'\, (\bar{\gamma}-\gamma_{t'}), \\
    & \dot{g}_{q, 2} + \gamma_t q \mathfrak{n}_{q, t} + g_{q, 2} q \Bigl( \frac{1}{2}\gamma_t - i \bar{\Lambda}_{q}^{-} + i\bar{\Lambda}_{q}^{+} \Bigr) = 0,  \\
    & \dot{g}_{q, 3} + \gamma_t q \mathfrak{n}'_{q, t} + g_{q, 3} q \Bigl( \frac{1}{2}\gamma_t + i\bar{\Lambda}_{q}^{-} - i \bar{\Lambda}_{q}^{+} \Bigr) = 0, \\
    & \dot{g}_{q, 4} + \gamma_t q \mathfrak{m}_{q, t} + g_{q, 4} q \Bigl( \frac{1}{2}\gamma_t + i\bar{\Lambda}_{q}^{-} + i\bar{\Lambda}_{q}^{+} \Bigr) = 0, \\
    & \dot{g}_{q, 5} + \gamma_t q \mathfrak{m}'_{q, t} + g_{q, 5} q \Bigl( \frac{1}{2}\gamma_t - i\bar{\Lambda}_{q}^{-} - i\bar{\Lambda}_{q}^{+} \Bigr) = 0. \\
\end{aligned}
\right.
\end{equation}
The resulting effective Liouvillian reads
\begin{equation}
\tilde{\mathcal{L}}
= \sum_{q>0}
  \biggl[ \frac{q}{2} \bigl( \bar{\Lambda}_{q}^{+} + \bar{\Lambda}_{q}^{-} \bigr) H_{q, 0} + iq \bigl( \bar{\Lambda}_{q}^{+} - \bar{\Lambda}_{q}^{-} \bigr) (D_{q, 1} + \bar{D}_{q, 2}) + \frac{\bar{\gamma}}{2} q (D_{q, 1} - \bar{D}_{q, 2}) \biggr].
\end{equation}
After vectorization as in Appendix~\ref{app:HOspectrum}, in order to ensure convergence of the $\rho = \sum_{n, m} \rho_{n, m}|m\rangle \otimes |n\rangle$ solving the eigenvalue problem $\tilde{\mathcal{L}} \rho = E \rho$, one can show that  $\tilde{\mathcal{L}}$ cannot contain terms with $D_{q, 2}$ or $\bar{D}_{q, 2}$.
To realize this, we perform a final time-independent rotation using $\mathcal{T} = \prod_{q>0} \mathcal{T}_q = \prod_{q>0} e^{\zeta v_{q, 3}}$ for some suitable complex number $\zeta$:
\begin{equation}
\begin{aligned}
\tilde{\mathcal{L}}
\mapsto
\mathcal{T} \tilde{\mathcal{L}} \mathcal{T}^{-1} 
& = \sum_{q>0} q \biggl[ \bar{\Lambda}_{q}^{-} e_{q, 1} + \bar{\Lambda}_{q}^{+} e_{q, 4} + \frac{\bar{\gamma}}{2} v_{q, 1} + \zeta \biggl( \frac{\bar{\gamma}}{2} - i \bigl( \bar{\Lambda}_{q}^{+} - \bar{\Lambda}_{q}^{-} \bigr) \biggr) v_{q, 3} \biggr] \\
& = \sum_{q>0} q
    \biggl[ \frac{1}{2} \bigl( \bar{\Lambda}_{q}^{+} + \bar{\Lambda}_{q}^{-} \bigr) H_{q, 0}
    + \biggl( \frac{\bar{\gamma}}{2} + i \bigl( \bar{\Lambda}_{q}^+ - \bar{\Lambda}_{q}^- \bigr) \biggr) D_{q, 1}
    + \zeta \biggl( \frac{\bar{\gamma}}{2} - i (\bar{\Lambda}_{q}^{+} - \bar{\Lambda}_{q}^{-}) \biggr) \bar{D}_{q, 1} \\
& \qquad \qquad + (\zeta-1) \biggl( \frac{\bar{\gamma}}{2} - i \bigl( \bar{\Lambda}_{q}^{+} - \bar{\Lambda}_{q}^{-} \bigr) \biggr) \bar{D}_{q, 2} \biggr].
\end{aligned}
\end{equation}
To eliminate the coefficient in front of $\bar{D}_{q, 2}$, we pick $\zeta = 1$, yielding the effective Liouvillian \eqref{eq:timeindepliouvchiral}.

\end{document}